\documentclass[a4paper,twocolumn,11pt, unpublished]{quantumarticle}
\pdfoutput=1
\usepackage[english]{babel}
\usepackage[utf8]{inputenc}
\usepackage{amsthm}
\usepackage{amssymb}
\usepackage{mathtools}
\usepackage{physics}
\usepackage{siunitx}
\usepackage{xcolor}
\usepackage{graphicx}
\usepackage{tabularx,ragged2e,booktabs,caption}
\newcolumntype{C}[1]{>{\Centering}m{#1}}

\usepackage{soul} 
\usepackage{svg}
\usepackage[left=23mm,right=13mm,top=35mm,columnsep=15pt]{geometry} 
\usepackage{comment}
\usepackage{adjustbox}
\usepackage{placeins}
\usepackage[T1]{fontenc}
\usepackage{lipsum}
\usepackage{csquotes}
\graphicspath{ {./Images/} }
\usepackage[pdftex, pdftitle={Article}, pdfauthor={Author}]{hyperref}

\usepackage[numbers,sort&compress]{natbib}

\begin{document}

\title{Quantum manipulation of a two-level mechanical system}

\author{Salvatore Chiavazzo}
    \email[Correspondence email address: ]{sc1026@exeter.ac.uk}
    \affiliation{Department of Physics and Astronomy, University of Exeter, Exeter, Devon EX4 4QL, UK}

\author{Anders S. Sørensen}
    \affiliation{Center for Hybrid Quantum Networks (Hy-Q), Niels Bohr Institute,
University of Copenhagen, Blegdamsvej 17, DK-2100 Copenhagen Ø, Denmark}

\author{Oleksandr Kyriienko}
    \affiliation{Department of Physics and Astronomy, University of Exeter, Exeter, Devon EX4 4QL, UK}

\author{Luca Dellantonio}
    \affiliation{Department of Physics and Astronomy, University of Exeter, Exeter, Devon EX4 4QL, UK}
    \affiliation{Institute for Quantum Computing, University of Waterloo, Waterloo, Ontario, Canada, N2L 3G1}
    \affiliation{Department of Physics \& Astronomy, University of Waterloo, Waterloo, Ontario, Canada, N2L 3G1}


\begin{abstract}

We consider a nonlinearly coupled electromechanical system, and develop a quantitative theory for two-phonon cooling. In the presence of two-phonon cooling, the mechanical Hilbert space is effectively reduced to its ground and first excited states, allowing for quantum operations at the level of individual phonons and preparing nonclassical mechanical states with negative Wigner functions. We propose a scheme for performing arbitrary Bloch sphere rotations, and derive the fidelity in the specific case of a $\pi$-pulse. We characterise detrimental processes that reduce the coherence in the system, and demonstrate that our scheme can be implemented in state-of-the-art electromechanical devices.
\end{abstract}

\maketitle

\section{Introduction}\label{sec:Intro}

The study of mechanical systems in the quantum regime has seen tremendous advancements in the last decades \cite{MarquardtReview}. Improved design and quality of optical cavities \cite{Papp2014,Wu2020,Lai2014}, microwave resonators \cite{Reagor2013,Kudra2020}, and mechanical oscillators \cite{Tsaturyan2014,SN2014,Yu2014,Tsaturyan2017,MacCabe2019} have facilitated studies of radiation pressure effects in highly coherent regimes. These include mechanical ground state cooling \cite{verhagen2012quantum,Cooling1,Cooling2,AmCooling1,Cooling3,Cooling6}, ultra precise sensing \cite{zobenica2017integrated,Teufel2009,Sensing1,Sensing2,ExpNumb6}, generation of non-classical light and mechanical states \cite{SqLight1,SqLight2,SqLight3,Wollman2015,Sillanpaa2017,Paternostro2011,Vanner2013,Coelho2016,Milburn2016,kanarinaish2021nongaussian,Clarke_2018}, back action cancellation \cite{QB-EvasionNBI,QB-Evasion2,rossi2018measurement}, and detection of gravitational waves \cite{Abbott2016,VirgoGrav}. The vast majority of these achievements works in a linearized optomechanical regime, where phonons are coupled to photons through bilinear coupling terms, so-called Gaussian interactions. Despite these successes, non-Gaussian interactions are required to extend the range of opto- and electromechanical applications to the generation of non-classical states with negative Wigner functions \cite{Braunstein2005}. Some proposals achieve this goal by post-selecting measurement results \cite{Paternostro2011,Milburn2016,kanarinaish2021nongaussian}, but for several application it is desirable to have a deterministic protocol. 
%
%

While limitations posed by the mass of mechanical oscillators make studies of 
non-Gaussian effects challenging, recent advancements 
\cite{Marquardt2008Nature,Viennot2018,ma2020nonclassical,Sletten2019,brawley2016nonlinear,NonlinearOpto,Cattiaux2020,valimaa2021multiphonon} sparked interest in nonlinear optomechanical systems that can possibly achieve, e.g., quantum non-demolition (QND) measurement of the phonon number \cite{dellantonio2018quantum} and two-phonon cooling \cite{Nunnenkamp2010}. However, multi-mode interactions make non-Gaussian interactions difficult to exploit \cite{YanbeiChen,Yanay}, and render the realization of theoretical proposals \cite{Nunnenkamp2010,Hauer2018,MarquardtFollow,MarquardtPRL,AtPhon,pistolesi2020proposal,yang2019ground} highly challenging. 

It was recently shown \cite{dellantonio2018quantum} that these problems can be alleviated for carefully designed systems. Here, we consider a similar setup and demonstrate that one can realistically achieve two-phonon cooling \cite{Nunnenkamp2010}, which is a process that decreases motional quanta by annihilating two phonons at the time.
Contrary to the QND detection of the phonon number studied in Ref.~\cite{dellantonio2018quantum}, the two-phonon cooling is frequency selective and only talks to a single mechanical mode, making it more experimentally accessible. As a paradigmatic example of non-Gaussian operations, we describe how to create and manipulate a mechanical qubit, and use the two-phonon cooling to deterministically prepare states with negative Wigner functions. The performance of our protocol is characterized by a single parameter $\lambda$, which is uniquely determined by the experimental setup and, importantly, includes detrimental multi-mode contributions.

\begin{figure}[htbp]
    \includegraphics[width=1\linewidth, angle=0]{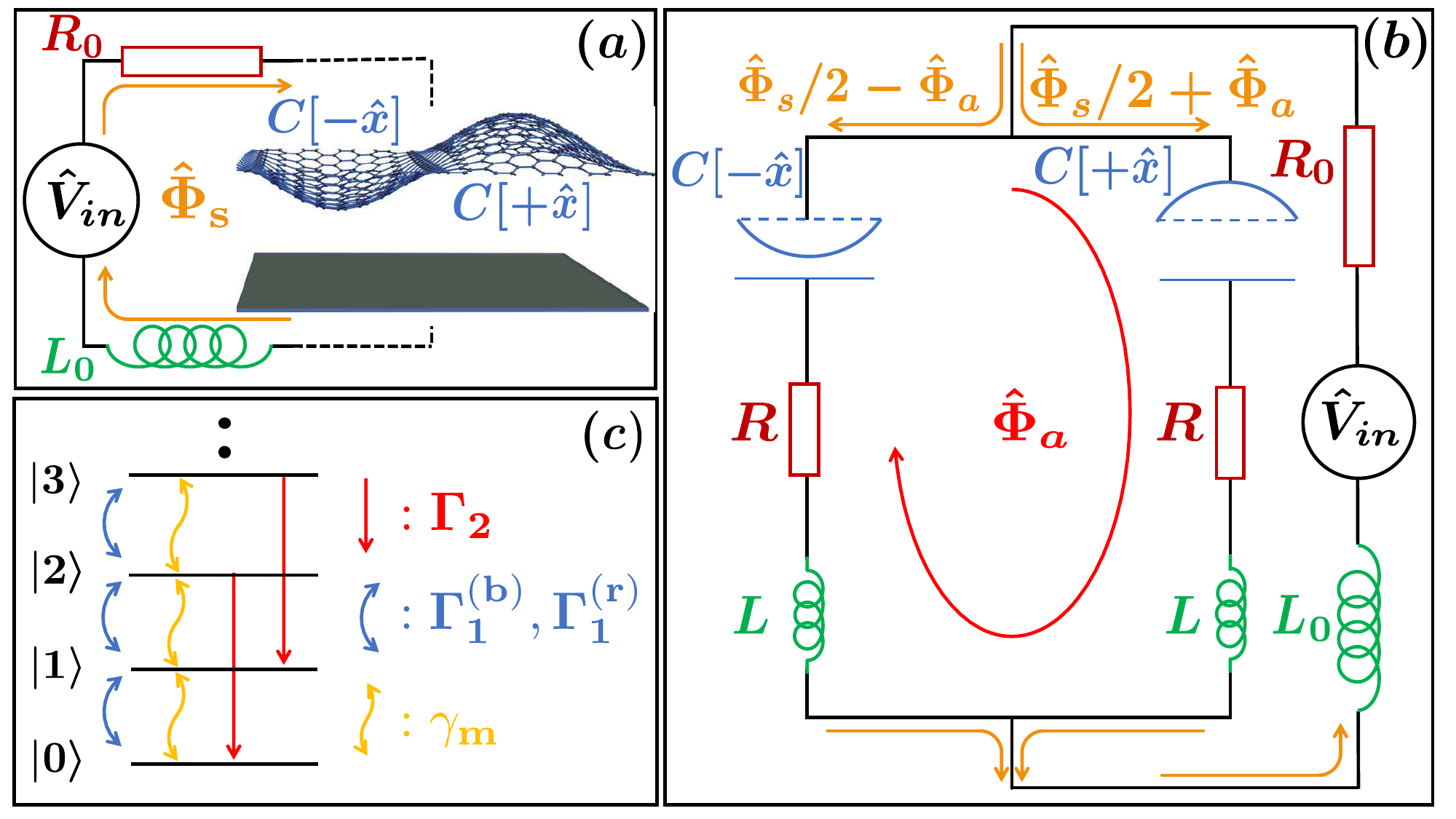}
    \caption{\textit{(a)}: A moving membrane forms one half of a capacitor in a driven $RLC$ circuit. The asymmetric oscillatory mode is chosen to suppress the linear coupling. \textit{(b)}: Electrical circuit considered in this work. The original capacitor is modelled by a loop containing two capacitors $C[\pm\hat{x}]$ (each representing half of the original one), resistances $R$ and inductances $L$. The loop describes redistribution of charges that generate a current proportional to the flux $\hat{\Phi}_{\rm a}$. The flux of the main electrical mode is $\hat{\Phi}_{\rm s}$, and flows from the resistor $R_0$ through the whole capacitor, and the inductance $L_{0}$. A generator with matching internal impedance (see App.~\ref{sec:mast_eq_der}) is used for driving $\hat{\Phi}_{\rm s}$ with an input field $\hat{V}_{\rm in}$. \textit{(c)}: Schematic of two-phonon cooling, with the rates from Eq.~\eqref{eq:master_equation}. Transitions between the membrane's Fock states $\lvert n \rangle$ are induced by the intrinsic mechanical reservoir (rate $\gamma_{\rm m}$), and the interactions with the electrical subsystems and their reservoirs (rates $\Gamma_2$, $\Gamma_{1}^{\rm (r)}$ and $\Gamma_{1}^{\rm (b)}$).
    }
    \label{fig:circuit}
\end{figure}

We study the $RLC$ circuit in Fig.~\ref{fig:circuit}(a), where one of the capacitor plates is a moving membrane. By changing the circuit's capacitance, its motion induces an electromechanical interaction. The considered asymmetric oscillatory mode ensures the suppression of the linear optomechanical coupling, making the higher-order (i.e. non-Gaussian) interactions dominant. The small magnitude of the quadratic coupling, however, obliges us to consider all possible residual linear effects that are detrimental to the two-phonon cooling. Following Ref.~\cite{dellantonio2018quantum}, we identify two malicious contributions. The first originates from fabrication imperfections, and couples the ``symmetric'' electrical mode (indicated by subscript ``s'' in Fig.~\ref{fig:circuit}) with the membrane. The second arises from redistribution of charges on the capacitor plates, which we model by introducing a loop in Fig.~\ref{fig:circuit}(b) with parasitic resistances $R$ and inductances $L$. The resulting ``asymmetric'' electrical mode (subscript ``a'') interacts linearly with the membrane. 

This work is organized as follows. In Sec.~\ref{sec:two_pho_cool}, we present the Master equation describing the system's dynamics and determine the conditions (summarized in the parameter $\lambda$) under which the two-phonon cooling is feasible. In Secs.~\ref{sec:expt_cons} and \ref{sec:qntm_man}, respectively, we discuss experimental implementations and introduce a protocol for coherently manipulating the membrane's quantum state within the reduced Hilbert space provided by the two-phonon cooling. Specifically, we consider a $\pi$ rotation of an arbitrary initial mechanical state and characterize its fidelity in terms of $\lambda$. We summarize our results and present an outlook for future works in Sec.~\ref{sec:conclusions}. 

%
\section{Two-phonon cooling} \label{sec:two_pho_cool}
Here, we derive the two-phonon cooling rate $\Gamma_2$ and compare it with the thermalization rates from detrimental linear effects. The optomechanical interaction described by the couplings $g_1$ and $g_2$ is derived by expanding the inverse capacitances of the two halves in Fig.~\ref{fig:circuit}(b) up to second order: $C^{-1}[ \pm\hat{x}] \simeq 1/C_0 \pm g_1 (1 \mp \delta) \hat{x}/(2 C_0 \omega_{\rm s}) + g_2 \hat{x}^2/(4 C_0 \omega_{\rm s})$. The sign uniquely identifies the capacitor in the circuit, as indicated in Figs.~\ref{fig:circuit}(a) and (b). The operator $\hat{x}$ is the mechanical displacement away from equilibrium,
and $\omega_{\rm s} \equiv [C_0(L + 2L_0)]^{-\frac{1}{2}}$, $\omega_{\rm a} \equiv (C_0 L)^{-\frac{1}{2}}$ are the frequencies of the two electrical modes, with $\omega_{\rm a} \gg \omega_{\rm s}$. Furthermore, $\delta$ takes into account the effects of nano-fabrication inaccuracies. Without these ($\delta = 0$), the sign difference in the expansion suppresses the linear interaction between the membrane and the symmetric mode.
The Hamiltonian of the system in Fig.~\ref{fig:circuit}(b) is $\hat{\mathcal{H}} = \hat{\mathcal{H}}_{\rm s} + \hat{\mathcal{H}}_{\rm a} + \hat{\mathcal{H}}_{\rm m}
+ \hat{\mathcal{H}}_{\rm int}$, where $\hat{\mathcal{H}}_{\rm s} = \hat{\Phi}_{\rm s}^{2} / (2L_0 + L) + \hat{Q}_{\rm s}^{2} / 4C_0 - 2 \hat{Q}_{\rm s} \hat{V}_{\rm in}$, $\hat{\mathcal{H}}_{\rm a} = \hat{\Phi}_{\rm a}^{2} / (4 L) + \hat{Q}_{\rm a}^{2} / C_0$, $\hat{\mathcal{H}}_{\rm m} = \hbar \omega_{\rm m} \hat{b}^{\dagger}\hat{b}$
, and 
\begin{equation}\label{eq:Hamiltonian}
\begin{split}
\hat{\mathcal{H}}_{\rm int} = & \delta g_{1}\frac{\hat{Q}_{\rm s}^{2}(\hat{b} + \hat{b}^{\dagger})}{4 C_{0} \omega_{\rm s}} + g_{1}\frac{\hat{Q}_{\rm a}\hat{Q}_{\rm s}(\hat{b} + \hat{b}^{\dagger})}{C_{0} \omega_{\rm s}} \\ & + g_{2}\frac{\hat{Q}_{\rm s}^2(\hat{b}\hat{b} + \hat{b}^{\dagger}\hat{b}^{\dagger})}{8 C_{0} \omega_{\rm s}}.
\end{split}
\end{equation}
Here, $\hat{Q}_{\mu}$ and $\hat{\Phi}_{\mu}$ correspond, respectively, to the charge sum ($\mu = {\rm s}$) or difference ($\mu = {\rm a}$) of the two capacitors, and the magnetic fluxes flowing in the circuit. The mechanical annihilation (creation) operator is $\hat{b}$ ($\hat{b}^{\dagger}$), and the membrane frequency is $\omega_{\rm m}$. The remaining parameters are shown in Fig.~\ref{fig:circuit}(b) and described in its caption. 

The interaction Hamiltonian in Eq.~\eqref{eq:Hamiltonian} is derived for the symmetric mode being driven with a strong field of frequency  $\omega_{\rm s} - 2 \omega_{\rm m}$. This allows disregarding all membrane modes except for the chosen asymmetric one. With this driving, the only resonant process is $\propto g_{2} \hat{Q}_{\rm s}^2 \hat{b}\hat{b}$, where a single photon from the drive and two phonons are converted into an excitation of the symmetric electrical mode. Using the rotating wave approximation, we thus neglect a term $\propto g_2 \hat{Q}_{\rm s}^2 \hat b^\dagger \hat b$, as well as the dynamical contribution from the quadratic coupling between the asymmetric electrical and mechanical modes. The latter follows from the absence of resonance with the asymmetric electrical mode, which is satisfied whenever $L_0 \gg L$. Finally, we have retained two linear terms $ \propto \delta g_{1} \hat{Q}_{\rm s}^{2}(\hat{b} + \hat{b}^{\dagger})$ and $\propto g_{1} \hat{Q}_{\rm a}\hat{Q}_{\rm s}(\hat{b} + \hat{b}^{\dagger})$ in Eq.~\eqref{eq:Hamiltonian} that are responsible for the detrimental heating processes described in the introduction. 

From the Hamiltonian, we derive the master equation for the mechanical system (see App.~\ref{sec:mast_eq_der} for details):
\begin{widetext}
\begin{equation}\label{eq:master_equation}
	\begin{aligned}
		\dot{\hat{\rho}}_{\rm m} =& \frac{1}{i \hbar} \left[ \mathcal{\hat{H}}_{\rm m}, \hat{\rho}_{\rm m} \right] + \frac{\gamma_{\rm m}}{2} \left( \bar{n}_{\rm m} + 1\right)  \mathcal{L}[\hat{b};\hat{\rho}_{\rm m}] 
		+
		\frac{\gamma_{\rm m}}{2} \bar{n}_{\rm m} \mathcal{L}[\hat{b}^\dagger;\hat{\rho}_{\rm m}] 
		+
		\frac{\Gamma_2}{4} \mathcal{L}[\hat{b} \hat{b};\hat{\rho}_{\rm m}] 
		\\
		& 
		+ 
		\frac{\Gamma_1^{\rm (r)} + \Gamma_1^{\rm (b)}}{2} \mathcal{L}[\hat{b};\hat{\rho}_{\rm m}] + \frac{1}{2} \left(
		\frac{\Gamma_1^{\rm (r)}}{9} + \Gamma_1^{\rm (b)} \right)
		\mathcal{L}[\hat{b}^\dagger;\hat{\rho}_{\rm m}] ,
	\end{aligned}
\end{equation}
\end{widetext}
where $\hat{\rho}_{\rm m}$ is the membrane's density matrix, and $\mathcal{L}[\hat{O}; \hat{\rho}_{\rm m}] = 2 \hat{O} \hat{\rho}_{\rm m} \hat{O}^\dagger - \acomm{\hat{O}^\dagger \hat{O}}{\hat{\rho}_{\rm m}}$ for any operator $\hat{O}$. In deriving Eq.~\eqref{eq:master_equation} we have assumed that the mechanical, the symmetric, and the asymmetric electrical modes are in contact with independent Markovian reservoirs with $\lbrace \text{decay rate, average occupation}\rbrace$ given by $\lbrace \gamma_{\rm m},\bar{n}_{\rm m}\rbrace$, $\lbrace \gamma \equiv R_0/(L_0 +L/2),0\rbrace$, and $\lbrace R/L,0 \rbrace$, respectively.
As shown in Fig.~\ref{fig:circuit}(c), $\Gamma_{2} = g_2^{2} \lvert \alpha \rvert^2 / 4\gamma$ is the two-phonon cooling rate, while $\Gamma_1^{\rm (r)} =  (\delta g_1)^{2} \lvert \alpha \rvert^2 \gamma/(2\omega_{\rm m}^2)$ and $\Gamma_1^{\rm (b)} =  g_{1}^{2} \lvert \alpha \rvert^2 \gamma R/(2 \omega_{\rm s}^2 R_0)$ correspond to unwanted linear processes. Here, $\lvert \alpha \rvert^2$ is the number of photons generated by the drive $\hat{V}_{\rm in}$ in the symmetric mode (see Fig.~\ref{fig:circuit}).

Eq.~\eqref{eq:master_equation} is valid in the weak coupling regime, where excitations transferred from the mechanical to the electrical subsystems cannot be transferred back coherently. This occurs when $\Gamma_{1}^{\rm (b)},\Gamma_{1}^{\rm (r)}, \Gamma_2 \ll \gamma$. Furthermore, we have assumed the resolved sideband regime $\gamma \ll \omega_{\rm m}$ to write the rates as above and neglect the sideband at frequency $\omega_{\rm s} + 2 \omega_{\rm m}$, associated with two-phonon heating. 

The relevant processes in the system are schematically represented in Fig.~\ref{fig:circuit}(c). The two-phonon cooling $\propto \Gamma_{2}$ arises from the quadratic term in Eq.~\eqref{eq:Hamiltonian}, while ``standard'' heating and cooling of a system coupled to a Markovian reservoir is described by the rates $\Gamma_{1}^{\rm (r)}$ and $\Gamma_{1}^{\rm (b)}$.
The first originates from the residual linear coupling $\delta g_{1}$ between the symmetric and the mechanical modes. The latter originates from cross-coupling between symmetric and anti-symmetric modes described by the three-body term $\propto g_{1}$ in Eq.~\eqref{eq:Hamiltonian}. In these processes, a single phonon can either be annihilated or created by interacting with a photon. The membrane is then both cooled down at a rate $\propto \Gamma_{1}^{\rm (r)} + \Gamma_{1}^{\rm (b)}$ and heated up at a rate $\propto \Gamma_{1}^{\rm (r)}/9 + \Gamma_{1}^{\rm (b)}$. The difference arises because the sideband at frequency $\omega_{\rm s} - \omega_{\rm m}$ is enhanced by being closer to resonance compared to the sideband at frequency $\omega_{\rm s} + \omega_{\rm m}$. The asymmetric mode, however, is assumed to be far detuned $\omega_{\rm a} \gg \omega_{\rm s}$, and therefore the associated heating and cooling rates $\Gamma_{1}^{\rm (b)}$ are the same. 

We note that both the desired two-photon cooling $\Gamma_2$ and the undesired heating processes $\Gamma_1^{\rm (r)}$, $\Gamma_1^{\rm (b)}$, are proportional to $\lvert \alpha \rvert^2$ and can be controlled by the strength of the drive. To understand whether the two-phonon cooling is dominant we define $\lambda = (\lambda_{\rm r}^{-1} + \lambda_{\rm m}^{-1})^{-1}$, where
\begin{subequations}\label{eq:lambda}
\begin{align}
\lambda_{\rm r} \equiv & \frac{\Gamma_{2}}{\Gamma_{1}^{\rm (r)}} = \frac{1}{2} \left(\frac{1}{\delta}\right)^2 \left( \frac{g_2}{g_1} \right)^2 \left( \frac{\omega_{\rm m}}{\gamma} \right)^2 , \label{eq:lambdaR}
\\
\lambda_{\rm m} \equiv& \frac{\Gamma_{2}}{\Gamma_{1}^{\rm (b)}} = \frac{1}{2} \left(\frac{g_2}{g_1} \right)^2 \left(\frac{\omega_{\rm s}}{ \gamma}\right)^2 \frac{R_0}{R} . \label{eq:lambdaB}
\end{align}
\end{subequations}
When the parameter $\lambda_{\rm r}$ ($\lambda_{\rm m}$) is larger than one, the non-Gaussian dynamics from the quadratic coupling prevails over the single phonon processes induced by the residual linear coupling $\delta g_{1}$ (multi-mode term proportional to $g_{1}$). Therefore, when $\lambda \gg 1$, the two-phonon cooling is the dominant process and the membrane is effectively confined to the two lowest states, thus representing a qubit. This mechanical qubit's coherence time $\lambda / \Gamma_2$ corresponds to the timescale at which unwanted processes act. We remark that the intrinsic mechanical damping is also detrimental for the two-phonon cooling, but its rate is not enhanced by the drive $\lvert \alpha \rvert^2$. While it is in principle possible to speed up all other processes to make it negligible, the small magnitude of the couplings and limitations on the driving power set requirements on the mechanical $Q$ factor necessary to achieve such a situation (see Sec.~\ref{sec:expt_cons} for details).
\begin{figure}[htbp]
\centering
\includegraphics[width=1\linewidth]{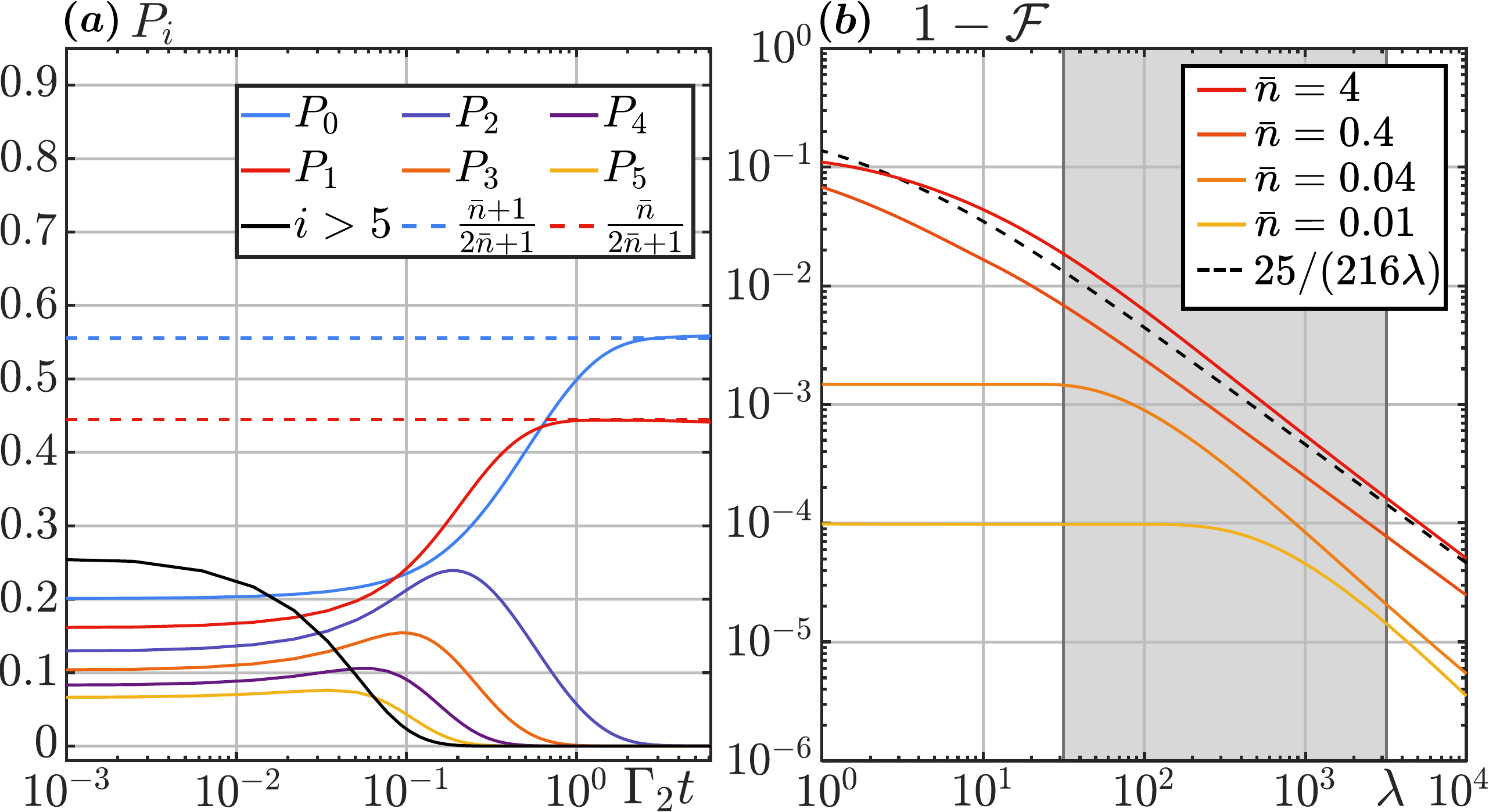}
    \caption{
    \textit{(a)}: Time evolution of the mechanical Fock states' populations $P_i = \langle i \rvert \hat{\rho}_{\rm m}\lvert i \rangle$ with two-phonon cooling active. We assume $\lambda = 2000$ and initialize the system in a thermal state with an average of $\bar{n} = 4$ phonons. The black line represents the sum $\sum_{j>5}\langle j \rvert \hat{\rho}_{\rm m}\lvert j \rangle$, and the other full lines represent the Fock states with $i\leq 4$. The  dashed lines are the expected populations of a  perfect two-phonon cooling process (see Sec.~\ref{sec:expt_cons}).
    \textit{(b)}: Minimum infidelity $1-\mathcal{F}$ as a function of $\lambda$ for different initial phonon averages $\bar{n}$. The dashed black line is the probability of the steady state being outside the $\lbrace \lvert 0 \rangle,\lvert 1 \rangle \rbrace$ subsector, and the shadowed area refers to realistic values of $\lambda$. In both plots, full curves are derived by simulating Eq.~\eqref{eq:master_equation} with the time chosen to maximize $\mathcal{F}$.
    }
\label{fig:plot_twopho_cool}
\end{figure}
\section{Experimental feasibility}\label{sec:expt_cons}

A realistic estimate of $\lambda$ can be derived from recent experiments \cite{Teufel2011sc,Cooling2,Cooling3,PhononQNDmw,Wollman2015,Sillanpaa2017,midolo2018nano,ExpNumb5,ExpNumb6}. We consider a rectangular monolayer graphene membrane with area $1\times0.3$ $\mu$m$^2$ oscillating at $\omega_{\rm m} = (2\pi)80$ MHz and suspended $d_0=10$~nm above a conducting plate, forming the capacitor [see Fig.~\ref{fig:circuit}(a)]. Following Refs. \cite{EmilThesis,dellantonio2018quantum}, we derive the ratio $g_2/g_1 = \pi^2 x_{\rm zpm}/(8d_0)$, where $x_{\rm zpm}\equiv \sqrt{\hbar/(2m \omega_{\rm m})}$ is the zero point motion's amplitude of a membrane of mass $m$. For the electrical circuit, we assume $\omega_{\rm s} = (2\pi)7$ GHz, $\gamma = (2\pi)150$ kHz and $\delta \in [10^{-3},10^{-2}]$, which leads to varying $\lambda$. To include the influence of both detrimental contributions we set $\Gamma_1^{\rm (b)} = \Gamma_1^{\rm (r)}$, implying $R/R_0 \in [0.8,0.008]$.
Under these assumptions we find $\lambda \in [34,3400]$. 

The number of photons in the cavity can be regulated to ensure that the system is in the weak coupling regime $\Gamma_2 \ll \gamma$. This condition is equivalent (once $\Gamma_2$ is plugged in) to $ N_{\rm mw} \ll 4\gamma^2/g_2^2 = 9 \times 10^{10} $. Here, $ N_{\rm mw}$ is the number of microwave photons in the cavity and we assumed $g_2 \sim (2\pi) 1$ Hz, corresponding to a stray capacitance $C_{\rm s} = 100 C_0$ (see below). The associated power $P_{\rm up}$ is calculated to be  $P_{\rm up} = \hbar \omega_{\rm s} N_{\rm mw} (4 \omega_{\rm m}^2 + \gamma^2)/\gamma = 448$ mW. The  input power $P_{\rm{in}}$ must therefore be upper bounded by $P_{\rm{in}} \ll P_{\rm{up}}$ in order to remain in the weak coupling limit. We remark that, being the input drive off-resonant, only a very small fraction of the power is dissipated into the device. The corresponding intra-cavity dissipated power associated with $P_{\rm up}$ found above is $P_{\rm up} \gamma^2/(4 \omega_{\rm m}^2) = 397$ nW.
Below,
we estimate that for $N_{\rm mw} \leq 10^{10}$, corresponding to an intra-cavity dissipated (input) power of $44$ nW ($50$ mW),
a mechanical quality factor 
$Q=\omega_{\rm m}/\gamma_{\rm m} \gtrsim 10^5$ is sufficient to be dominated by the driving-induced processes (see also App.~\ref{sec:optical_regimes}). Therefore, we set $\gamma_{\rm m}=0$ in the following and explain how to generalize our results to non-negligible values of $\gamma_{\rm m}$ at the end of this section.

To understand how the two-phonon cooling works, we simulate Eq.~\eqref{eq:master_equation} for the parameters above and an initial thermal state of $\bar{n}$ average phonons (for the evolution of an initial coherent state, see App.~\ref{sec:pureStatesEvolution}). Fig.~\ref{fig:plot_twopho_cool}(a) shows the time dependence of the mechanical Fock states' amplitudes for $\bar{n}=4$ and $\lambda = 2000$. From an initial thermal distribution, highly populated states decay into lower ones until the ground and first excited states remain. Horizontal dashed lines indicate the expected populations of perfect two-phonon cooling; $(\bar{n}+1)/(2\bar{n}+1)$ for $\lvert 0 \rangle$ and $\bar{n}/(2\bar{n}+1)$ for $\lvert 1 \rangle$. In Fig.~\ref{fig:plot_twopho_cool}(b), full lines represent the infidelity $1-\mathcal{F} = 1 - \Tr{\left(\sqrt{\hat{\rho}_{\rm exp}} \hat{\rho}_{\rm m} \sqrt{\hat{\rho}_{\rm exp}}\right)^{\frac{1}{2}}}^2$ \cite{miszczak2008sub, jozsa1994fidelity} between $\hat{\rho}_{\rm m}$ and this desired outcome $\hat{\rho}_{\rm exp} = \left[(\bar{n} + 1) \lvert 0 \rangle \langle 0 \rvert + \bar{n} \lvert 1 \rangle \langle 1 \rvert \right]/(2\bar{n}+1)$. 
For all values of the parameters $\lambda$ and $\bar{n}$, the simulation runs for the time maximizing $\mathcal{F}$. For longer times, the redistribution of population within the $\{ \vert 0 \rangle, \vert 1 \rangle \}$ subspace caused by linear processes and heating lowers the fidelity (see App.~\ref{sec:dynmc_cntrbtns} for details). 

As it is possible to see from Fig.~\ref{fig:plot_twopho_cool}(b), $\mathcal{F}$ scales as $1/\lambda$, reflecting the compromise between the two-phonon cooling and the detrimental processes described above. Since a longer time is required for decaying into the ground and first excited states, lower fidelities are reached for higher $\bar{n}$. However, the probability $\sim 25/(216 \lambda)$ of the steady state being outside the $\{ \vert 0 \rangle, \vert 1 \rangle \}$ subsector [dashed black line in Fig.~\ref{fig:plot_twopho_cool}(b)] is independent of the initial state, highlighting that the two-phonon cooling is always the dominant process. The shadowed area corresponds to the range $\lambda \in [34,3400]$  found above, demonstrating that non-Gaussian operations are within reach of current technology.

In the remainder of this section, we describe experimental limitations originating from a parasitic capacitance $C_{\rm s}$, from extra heating induced by intracavity photons, and from the mechanical rate $\gamma_{\rm m}$. Starting with the first, the effect of $C_{\rm s}$ is to limit the coupling strengths $g_1$ and $g_2$, which are proportional to $1/(C_0 + C_{\rm s})$. A stray capacitance that is hundreds or even thousands of times bigger than $C_0$ is expected in realistic scenarios \cite{ExpNumb1,ExpNumb5,ExpNumb6,ExpNumb7,Sensing3,Sensing4}. Assuming $C_{\rm s} = 100 C_0$ \cite{ExpNumb1}, the linear coupling for the example parameters considered above becomes $g_1\sim (2\pi)7$ kHz, and the quadratic $g_2 \sim (2\pi)1$ Hz.  Since the optomechanical couplings enter in $\lambda$ as a ratio, stray capacitances do not have a direct effect on the realization of the two-phonon cooling. In the model considered so far, both a limited value of the $Q$ factor and the presence of a stray capacitance can thus be compensated with a stronger drive. 

In realistic examples, however, one cannot simply increase the driving strengths to arbitrary values to neglect the effects of the environment. Whether the system can tolerate high powers is a complex question which depends on the details of the experimental setup. As explained above, the photon flux that needs to be sent into the circuit to overcome the intrinsic heating depends on the mechanical quality factor $Q$ and the absolute value of the couplings. 
With $C_{\rm s} = 100 C_0$ \cite{ExpNumb1} and $Q \in [10^5,10^7]$, the required number of microwave photons $N_{\rm mw}$ in the cavity can vary from $10^{10}$ to $10^8$ \cite{dellantonio2018quantum}. The question is then how much heating is induced by these photons. As an example, in Ref.~\cite{TemperatureRising} heating effects are observed starting from $N_{\rm mw} \gtrsim 10^8$. We believe, however, that compared to that work the heating may be reduced for our system. The experimental results indicate that the heating is due to electric fluctuations in the main electrical mode $\hat{\Phi}_{\rm s}$.  This heating should be reduced by the vanishing linear coupling $\delta g_1$ (due to symmetry).

 We may now determine the contribution from $\gamma_{\rm m}$ to $\lambda$. Since $\lambda$ is defined as the ratio between the two-phonon cooling rate $\Gamma_2$ and the combined rate at which the system's state looses its coherence, it is possible to include $\gamma_{\rm m}$ into $\lambda$ by redefining
\begin{equation}
    \lambda' = \lambda \frac{\Gamma_1^{\rm(b)}+\Gamma_1^{\rm(r)}}{\Gamma_1^{\rm(b)}+\Gamma_1^{\rm(r)}+(2\bar{n}_{\rm m} + 1) \gamma_{\rm m}}.
\end{equation}
All conclusions drawn for $\lambda$ above and in the following Sec.~\ref{sec:qntm_man}, hold upon the substitution $\lambda \rightarrow \lambda'$. Therefore, whether or not the mechanical damping $\gamma_{\rm m}$ can be ignored depends on the ratio $[\Gamma_{1}^{\rm (b)} + \Gamma_{1}^{\rm (r)}]/[(2\bar{n}_{\rm m}+1)\gamma_{\rm m}]$. By increasing the flux of photons $\lvert \alpha\rvert^2$ (and consequently $N_{\rm mw}$), the numerator is enhanced, which reduces the  mechanical quality factor $Q$ required for neglecting the intrinsic damping $\gamma_{\rm m}$.

While several aspects must be considered for realistic experimental scenarios, the flexibility of the system allows overcoming imperfections such as stray capacitances and intrinsic heating.
Two detrimental contributions to the desired system dynamics that are challenging to be estimated are mechanical dephasing and additional heating effects coming from the driving. As commented above, there are qualitative arguments suggesting that the latter
does not jeopardize the two-phonon cooling. Dephasing, on the other side, is expected to be less or at most comparable to the intrinsic decay rate $\gamma_{\rm m}$ \cite{Mirhosseini2019, Wilson2015}. However, by its nature the two-phonon cooling is resilient against
it,
making dephasing detrimental only for the coherent manipulation of the mechanical qubit -- see the following Sec.~\ref{sec:qntm_man}. Importantly, even in scenarios in which the dephasing may be the dominant decoherence rate in the system, our proposal allows for implementing qubit rotations (see below) and thereby spin echo techniques \cite{Hahn1950} that can mitigate its detrimental effects.

\section{Quantum manipulation}\label{sec:qntm_man} 

Above, we considered the two-phonon cooling and the condition to reduce the membrane's infinite dimensional Hilbert space onto a two level system. Here, we introduce a protocol to perform arbitrary rotations on the Bloch sphere of the resulting mechanical state.
We assume the system to be initialized in a suitable state and later read out using e.g., Gaussian operations. 
Initialization can be done via ground state cooling \cite{yi2014ground,Cooling1, Cooling2, verhagen2012quantum} (a technique that can be implemented in the parameter regimes considered in Sec.~\ref{sec:expt_cons} using the linear coupling introduced below), or by collapsing the wave function via detection. The latter requires $\lambda \gg 1$ \cite{dellantonio2018quantum}, similarly to the two-phonon cooling.

\begin{figure}[thbp]
    \centering
    \includegraphics[width = 0.5 \textwidth]{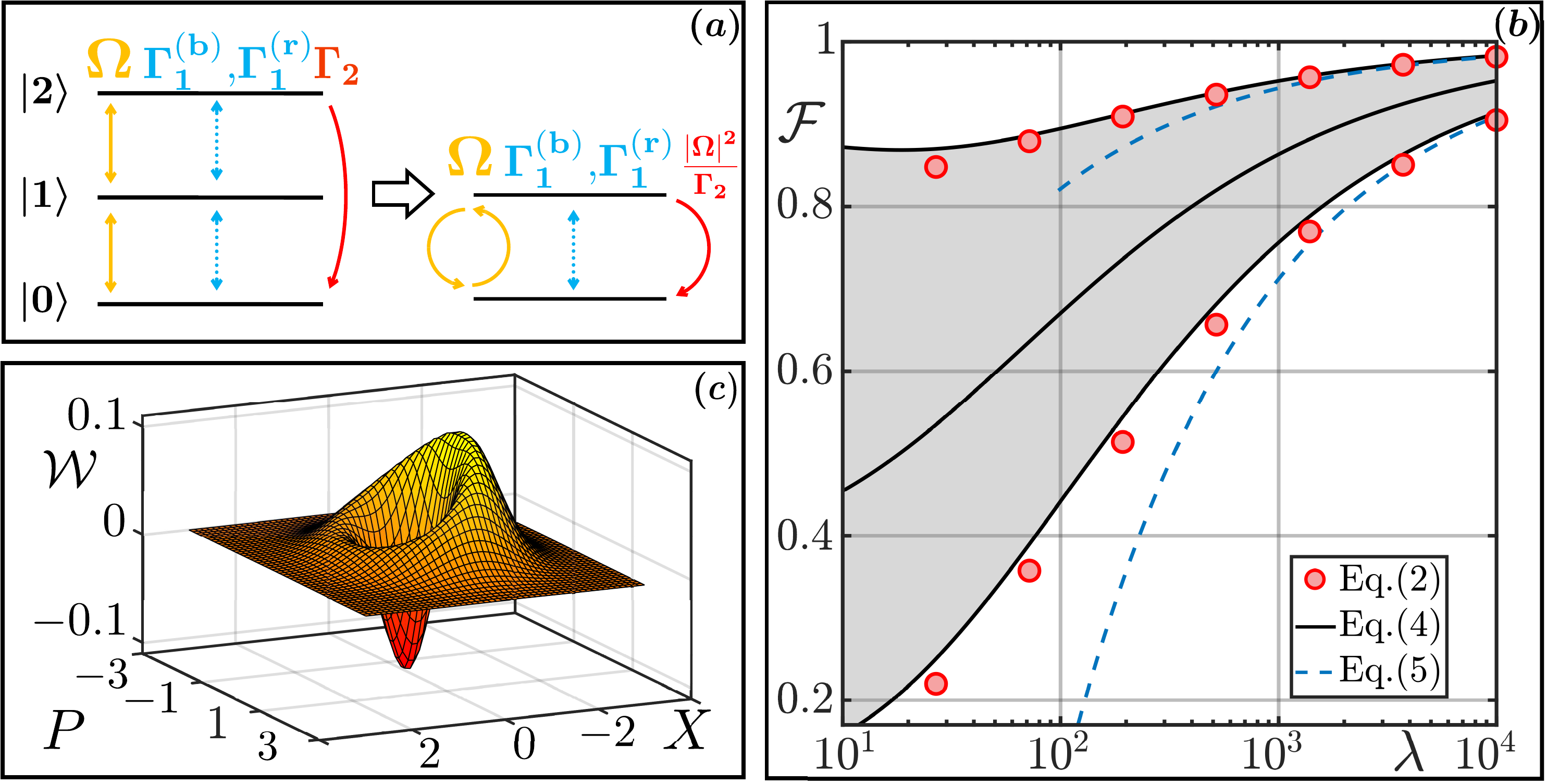}
    \caption{\textit{(a)}: Levels, couplings and decay rates considered for the rotation of the mechanical Bloch sphere. As explained in the text, $\lvert 2 \rangle$ is adiabatically eliminated, leading to the additional decay $\propto\lvert \Omega \rvert^{2}/\Gamma_{2}$ in an effective two level system. \textit{(b)}: Fidelity of a $\pi$-pulse as a function of $\lambda$. Full lines indicate the worst, the best, and the average fidelities -- depending on the initial conditions. The shadowed area comprises the values which can be reached according to Eqs.~\eqref{sys:rhodot_2times2}, and the dots are derived by numerically simulating Eq.~\eqref{eq:master_equation}, with the same initial conditions as for the neighbouring black line. Dashed blue curves are the asymptotic expansions from Eq.~\eqref{eq:fidelity}.
    \textit{(c)}: Wigner function $\mathcal{W}$ of the mechanical state after a $\pi$-pulse with initial state $\lvert 0 \rangle$ for $\lambda = 20$. $X$ and $P$ are the dimensionless 
    position and momentum operators, respectively. For all plots we assume $\Gamma_{1}^{\rm (b)} = \Gamma_{1}^{\rm (r)}$ and $\gamma_{\rm m}=0$.}
    \label{fig:quadrant}
\end{figure}

We assume that the two-phonon cooling is constantly running and is dominant compared to $\Gamma_{1}^{\rm (r)}$ and $\Gamma_{1}^{\rm (b)}$, i.e., $\lambda \gg 1$. Furthermore, as sketched in Fig.~\ref{fig:quadrant}(a), we modify the mechanical Hamiltonian to include an asymmetric linear drive: $\hat{\mathcal{H}}_{\rm m} \rightarrow \hat{\mathcal{H}}_{\rm m} + \hbar ( \Omega\hat{b} +  \Omega^{*}\hat{b}^\dagger)/2$. This driving term can be obtained by introducing two additional input fields that are off-resonant with the previously studied system dynamics and whose frequency difference equals $\omega_{\rm m}$ (for additional informations, see App.~\ref{sec:drivingField}). By changing their strengths and phases, it is possible to control $\Omega$, which determines the rotation on the Bloch sphere. We choose $\Omega$ such that $\Gamma_{1}^{\rm (r)},\Gamma_{1}^{\rm (b)} \ll \lvert \Omega \rvert \ll \Gamma_{2}$. 
These conditions, which can always be satisfied when $\lambda \gg 1$, ensure that the dynamics is constrained to the two lowest energy levels (despite the linear drive coupling the system to the second excited state) and that the coherent drive is dominant compared to the undesired thermalization processes.

Although Eq.~\eqref{eq:master_equation} can be used to compute the exact time evolution of the system, we derive an approximate solution to grasp the essential physics. We truncate the mechanical Hilbert space to the ground ($\lvert 0 \rangle$) and first two excited ($\lvert 1 \rangle$, $\lvert 2 \rangle$) states.
Since $\Gamma_{2} \gg \lbrace \lvert \Omega \rvert, \Gamma_1^{\rm (b)}, \Gamma_1^{\rm (r)}\rbrace$, $\lvert 2 \rangle$ can be adiabatically eliminated [see Fig.~\ref{fig:quadrant}(a)] and we get 
\begin{subequations}\label{sys:rhodot_2times2}
	\begin{align}
	\begin{split}
	\dot{\hat{\rho}}_{x} =
	& - 2\left( \frac{\lvert\Omega\rvert^2}{4\Gamma_2} +  \frac{\Gamma^{\rm (r)}_1}{3} +  \Gamma_1^{\rm (b)}\right) \hat{\rho}_{x}
	\\&
	- \Im{\Omega} (1 - 2 \hat{\rho}_{11}),
	\end{split}
	\\
	\begin{split}
	\dot{\hat{\rho}}_{y} =&   - 2\left( \frac{\lvert\Omega\rvert^2}{4\Gamma_2} +  \frac{\Gamma^{\rm (r)}_1}{3} +  \Gamma_1^{\rm (b)}\right) \hat{\rho}_{y}
	\\&
	+ \Re{\Omega} (1 - 2 \hat{\rho}_{11}) ,
	\end{split}
	\\
	\begin{split}
	\dot{\hat{\rho}}_{11} =& - 4\left( \frac{\lvert\Omega\rvert^2}{4\Gamma_2} +  \frac{\Gamma^{\rm (r)}_1}{3} +  \Gamma_1^{\rm (b)}\right) \hat{\rho}_{11} + \Gamma_1^{\rm (b)}
	\\&
	+ \frac{\Gamma_1^{\rm (r)}}{9} + 2 \left( \Re{\Omega} \hat{\rho}_{y}  - \Im{\Omega} \hat \rho_{x} \right),
	\end{split}
	\end{align}
\end{subequations}
where $\hat{\rho}_{ij}$ is the $\{i,j\}$ component of the density matrix $\hat{\rho}_{\rm m}$, $\hat{\rho}_{x} = \hat{\rho}_{10} + \hat{\rho}_{01}$, $\hat{\rho}_{y} = i (\hat{\rho}_{10} - \hat{\rho}_{01})$, and $\hat{\rho}_{00} = 1 - \hat{\rho}_{11}$. 
To characterize the rotation, we consider a $\pi$-pulse applied to an arbitrary state $\eta \lvert 0 \rangle + \beta \lvert 1 \rangle$. The drive $\Omega$ is activated for the time $\pi/\lvert\Omega\rvert$ required to switch $\lvert 0 \rangle$ into $\lvert 1 \rangle$ and vice versa. The quality of the evolution depends on the initial state and $\lambda$. By expanding the fidelity $\mathcal{F}$ in terms of $\lambda$, we find
\begin{widetext}
\begin{equation}
\small
\label{eq:fidelity}
\begin{split}
   \mathcal{F} [\lambda, \eta, \beta] \simeq 1 - \frac{1}{\sqrt{\lambda}} \left[ \left(1 + \frac{\Gamma_{1}^{\rm (b)}}{\Gamma_{1}^{\rm (r)}} \right) \left(1 + 3\frac{\Gamma_{1}^{\rm (b)}}{\Gamma_{1}^{\rm (r)}} \right) \right]^{-\frac{1}{2}} 
   \times & 
   \left[ \frac{\sqrt{3}\pi}{2}\left(1 + 3\frac{\Gamma_{1}^{\rm (b)}}{\Gamma_{1}^{\rm (r)}} \right) - \frac{2\Im{\eta \beta}}{3\sqrt{3}}\left(11 + 27\frac{\Gamma_{1}^{\rm (b)}}{\Gamma_{1}^{\rm (r)}} \right) \right. \\
   & \left.
   \qquad - \frac{2 \pi\Re{\eta \beta}^{2}}{\sqrt{3}}\left(1 + 3\frac{\Gamma_{1}^{\rm (b)}}{\Gamma_{1}^{\rm (r)}} \right) \right],
\end{split}
\end{equation}
\end{widetext}
where $\Omega = 2 \sqrt{(\Gamma_{1}^{\rm (r)}/3 + \Gamma_{1}^{\rm (b)})\Gamma_{2}}$ is chosen to maximize $\mathcal{F}$. This value of $\Omega$ reflects the compromise between exceeding the thermalization rates $\Gamma_{1}^{\rm (r)}$ and $\Gamma_{1}^{\rm (b)}$, and having a sufficiently low error rate $\lvert \Omega\rvert^2/(4\Gamma_2)$ from driving level $\lvert 2 \rangle$.

In Fig.~\ref{fig:quadrant}(b), we plot the fidelity of the $\pi$-pulse against $\lambda$ for $\Gamma_{1}^{\rm (r)} = \Gamma_{1}^{\rm (b)} = \Gamma_{1}$ and $\Omega = 4\sqrt{\Gamma_{1}\Gamma_{2}/3}$. Full and dashed lines are derived from Eqs.~\eqref{sys:rhodot_2times2} and \eqref{eq:fidelity}, respectively, and indicate minimum, maximum and average\footnote{The average is calculated assuming a uniform distribution of states over the Bloch sphere.} $\mathcal{F}$ with respect to $\eta$ and $\beta$. Dots are obtained by numerically simulating Eq.~\eqref{eq:master_equation} with the same parameters as for the corresponding full curves. Agreement is excellent for $\lambda \gtrsim 10$, indicating that the approximations used in deriving Eqs.~\eqref{sys:rhodot_2times2} are satisfied. The expression in Eq.~\eqref{eq:fidelity}, representing the asymptotic behaviour $\propto 1/\sqrt{\lambda}$, is valid for $\lambda \gtrsim 10^3$ and gives a lower bound for the real fidelity.

As shown in Fig.~\ref{fig:quadrant}(b), the system performs near perfect non-Gaussian operations, provided $\lambda$ is sufficiently large. However, even for limited $\lambda$, it is possible to deterministically produce interesting non-classical states. By applying the $\pi$-pulse to a system initially cooled to its ground state, we find that $\lambda \gtrsim 0.25$ is sufficient to show negativity of the Wigner function $\mathcal{W}$ (in contrast to the phonon number QND detection which requires $\lambda \gtrsim 50$ to see clear non-classical features \cite{dellantonio2018quantum}). Fig.~\ref{fig:quadrant}(c) demonstrates that highly negative values can be obtained for $\lambda = 20$. Despite it representing a mixture between $\lvert 0 \rangle$ and $\lvert 1 \rangle$, the negativity of the Wigner function confirms the non-classical properties of the state for much lower values of $\lambda$.
Specifically, for $\lambda=0.25$ the overlap $\mathcal{F}$ of the resulting state with the desired single phonon state is $\sim 56\%$, which is sufficient for negativity (Ref.~\cite{Sekatski2021} shows that the Wigner function is negative for any state with more than $50\%$ chance of being in the first excited state). For higher $\lambda$, $\mathcal{W}$ resembles more and more the one of a single phonon, and the overlap monotonically increases. As a reference, for $\lambda = 10^3$ we get $\mathcal{F} \sim 95\%$.


%
\section{Conclusions and outlook}\label{sec:conclusions}
We propose a protocol to realize non-Gaussian interactions in electromechanics.  This can considerably enlarge the class of states which can be prepared in mechanical systems, and thereby the phenomena that can be studied. Our setup relies on two-phonon cooling of the mechanical subsystem, enabled by the quadratic coupling between the membrane and the symmetric electrical mode. Interestingly, the conditions under which the two-phonon cooling is dominant are the same as for performing QND detection of the phonon number \cite{dellantonio2018quantum}. As opposed to the QND measurement, however, the two-phonon cooling is frequency selective, addresses only a single mode, and makes all others dynamically irrelevant. 

Even for modest values of $\lambda \gtrsim 1$, the two-phonon cooling allows for preparing states with negative Wigner functions. For large values of $\lambda$, the mechanical system can in principle be used for quantum information processing. 
The two-phonon cooling provides a non-Gaussian operation that, in combination with Gaussian interactions, allows for multi-mode complex operations \cite{Braunstein2005,LloydQC}. For instance, a bilinear interaction of the kind $\hat{b}_1\hat{b}^\dagger_2 + \hat{b}_2\hat{b}^\dagger_1$ (subscripts refer to individual mechanical modes), in combination with the two-phonon cooling, provides a $\sqrt{\rm SWAP}$ gate \cite{Schuch2003}. Along with the qubit rotations considered here, this constitutes a universal gate set, thus allowing for creating arbitrary multi-mode mechanical states, restricted within the two lowest energy levels of all oscillators.

While there are different ways to realize non-Gaussian interactions in mechanical systems, these generally require the strong coupling regime $g_1 \gg \gamma$ \cite{Marquardt2008Nature,YanbeiChen}, or resort to the interaction with additional quantum systems such as superconducting devices \cite{Sletten2019,valimaa2021multiphonon}. From one side, our scheme does not rely on additional quantum systems, making it simpler to realize and protecting it from additional detrimental effects coming from, e.g., these systems' thermal reservoirs. On the other side, the suppression of the linear coupling alongside the high tunability of electromechanical setups, allows for greatly lowering the constraints that must be satisfied to exploit non-Gaussian interactions in electromechanical systems (for details, see App.~\ref{sec:optical_regimes}). For these reasons, we expect this work to greatly facilitate the development of future devices that can overcome the limitations posed by Gaussian interactions.

\section*{Code Availability}
The code and nummerical simulations are available upon request.

\section*{Acknowledgements} \label{sec:acknowledgements}
S.\,C. thanks Christine Muschik for hosting at the University of Waterloo, where most of the research activity was conducted, and Jan F. Haase for valuable theoretical input. We thank Marco and Emanuele Grimaldo for fruitful discussions, and two anonymous Reviewers for valuable comments about the manuscript. This work has been supported by Transformative Quantum Technologies Program (CFREF), NSERC and the New Frontiers in Research Fund. S.\,C. and O.\,K. acknowledge the support from UK EPSRC New Investigator Award under the Agreement No. EP/
V00171X/1. A.\,S.\,S. acknowledges financial support by Danish National Research Foundation (Center of Excellence Hy-Q), and L.\,D. by the EPSRC quantum career development grant EP/ W028301/1.

\onecolumn\newpage
\appendix

\section{Derivation of the Master equation}\label{sec:mast_eq_der}

\bigskip
\begin{minipage}{\linewidth}
\centering
\captionof{table}{System parameters.} \label{tab:table1} 
\begin{tabularx}{\textwidth}{| c | X | c | X |}\toprule[1.25pt]
\hline
$\hat Q_{\mathrm{s}}$   & symmetric charge & $\hat \Phi_{\mathrm{s}}$ &  symmetric flux   \\ \hline
$\hat Q_{\mathrm{a}}$    & asymmetric charge & $\hat \Phi_{\mathrm{a}}$  & asymmetric flux   \\ \hline
$C_{\mathrm{0}}$    & system capacitance & $L_{\mathrm{0}}$  & system inductance \\ \hline
$\hat F_{\mathrm{m}}$   &  mechanical noise & $L$  & parasitic inductance \\ \hline
$x_{\mathrm{zpm}}$   &  mechanical zero point motion amplitude & $\omega_{\rm{in}}$  & driving frequency \\ \hline
$\hat V_{\mathrm{in}}$   & driving field & $\hat V_{\mathrm{R0}}$ & $R_0$ Nyquist noise   \\ \hline
$\hat V_{\mathrm{R1}}$   & $R_1$ Nyquist noise & $\hat V_{\mathrm{R2}}$ & $R_2$ Nyquist noise   \\ \hline
$\omega_{\mathrm{m}}$   &  mechanical resonant frequency & $\omega_{\mathrm{s}}$ & cavity resonant frequency  \\ \hline
$g_{\mathrm{1}}$   & linear coupling coefficient & $g_{\mathrm{2}}$ & nonlinear coupling coefficient  \\ \hline
$\delta g_{\mathrm{1}}$   & residual linear coupling & $\gamma_{\mathrm{m}}$ & mechanical decay rate  \\ \hline
$\gamma_{\mathrm{r}}$   & intrinsic decay rate of the symmetric mode & $\gamma_{\mathrm{t}}$ & decay rate of the symmetric mode into the transmission line \\ \hline
$\gamma$   & total decay rate of the symmetric mode & $\gamma_{\mathrm{l}}$ & total decay rate of the asymmetric mode  \\ \hline
\bottomrule[1.25pt]
\end{tabularx}
\par
\bigskip
\end{minipage}

In Sec.~\ref{sec:two_pho_cool} we use the master equation [Eq.~\eqref{eq:master_equation}] to describe the electromechanical circuit in Fig.~\ref{fig:circuit}(b). Here, we provide the details of its derivation. As discussed in Sec.~\ref{sec:two_pho_cool}, the full Hamiltonian $\hat{\mathcal{H}}$ of the system corresponds to the sum of four terms $\hat{\mathcal{H}} = \hat{\mathcal{H}}_{\rm m} + \hat{\mathcal{H}}_{\rm s} + \hat{\mathcal{H}}_{\rm a} + \hat{\mathcal{H}}_{\rm int}$, where
\begin{subequations}\label{eq:FullHamiltonian}
\begin{align}
    \hat{\mathcal{H}}_{\rm m} = & \hbar \omega_{\rm m} \hat{b}^{\dagger}\hat{b} - x_{\rm zpm} \hat{F}_{\rm m} (\hat{b} + \hat{b}^\dagger),
    \\
    \hat{\mathcal{H}}_{\rm s} = & \frac{\hat{Q}^2_{\rm s}}{4 C_{\rm 0}} + \frac{\hat{\Phi}^2_{\rm s}}{L + 2 L_{0}} - \hat{Q}_{\rm s} (2\hat{V}_{\rm in} + 2\hat{V}_{\rm R0} + \hat{V}_{ \rm R2}+ \hat{V}_{\rm R1} ),
    \\
    \hat{\mathcal{H}}_{\rm a} = & \frac{\hat{Q}^2_{\rm a}}{C_{\rm 0}} + \frac{\hat{\Phi}^2_{\rm a} }{4 L}- 2 \hat{Q}_{\rm a} (\hat{V}_{\rm R2} - \hat{V}_{\rm R1}),
    \\
    \begin{split}
    \hat{\mathcal{H}}_{\rm int} = & \frac{g_{\rm 1}}{C_{\rm 0} \omega_{\rm s}} \hat{Q}_{\rm a} \hat{Q}_{\rm s} (\hat{b} + \hat{b}^\dagger) +\frac{\delta g_{\rm 1} }{ \omega_{\rm s} C_{\rm 0}} \hat {Q}_{\rm a}^2 (\hat{b} + \hat{b}^\dagger) + \frac{\delta g_{\rm 1}}{4 \omega_{\rm s} C_{\rm 0}} \hat {Q}_{\rm s}^2 (\hat{b} + \hat{b}^\dagger) 
    \\
    & + \frac{g_{\rm 2} }{2 \omega_{\rm s} C_{\rm 0}} \hat {Q}_{\rm a}^2 (\hat{b} + \hat{b}^\dagger) ^2 + \frac{g_{\rm 2}}{8 \omega_{\rm s} C_{\rm 0}} \hat {Q}_{\rm s}^2 (\hat{b} + \hat{b}^\dagger) ^2.
    \end{split}
    \label{eq:FullHamiltonianInt}
\end{align}
\end{subequations}
The Hamiltonian $\hat{\mathcal{H}}$ describes the dynamics of three interacting subsystems: the mechanical membrane of frequency $\omega_{\rm m}$, the symmetric electrical mode of frequency $\omega_{\rm s} \equiv [C_0 (L+2L_0)]^{-\frac{1}{2}}$, and the asymmetric electrical mode of frequency $\omega_{\rm a} \equiv (C_0 L)^{- \frac{1}{2}}$. Throughout our work, we assume $\omega_{\rm m} \ll \omega_{\rm s} \ll \omega_{\rm a}$.
For completeness, in Eqs.~\eqref{eq:FullHamiltonian} we include all terms, such as the noise contributions, which were omitted in Eq.~\eqref{eq:Hamiltonian}.  
Specifically, $\hat{V}_{\rm R0}$, $\hat{V}_{\rm R1}$ and $\hat{V}_{\rm R2}$ are the Johnson-Nyquist \cite{landauer1989johnson} noises associated with the resistor $R_0$ and the two resistors $R$, respectively [see Fig.~\ref{fig:circuit}]. $\hat{F}_{\rm m}$, which is associated to the mechanical damping $\gamma_{\rm m}$, describes a random force acting on the membrane as a white noise. The additional terms that appear in Eq.~\eqref{eq:FullHamiltonianInt} but are absent from Eq.~\eqref{eq:Hamiltonian} arise from the expansions of the two halves of the capacitor $C[\pm \hat{x}]$ and are the ones proportional to $\delta g_1 \hat{Q}_{\rm a}^2$, $g_2 \hat{Q}_{\rm a}^2$, and $g_2\hat{Q}_{\rm s}^{2}\hat{b}^{\dagger}\hat{b}$ in Eq.~\eqref{eq:FullHamiltonianInt}. We explain in the following that they are negligible in the considered parameter regime.

Assuming a strongly driven coherent field $\hat V_{\rm in}$ at the frequency $\omega_{\rm in} = \omega_{\rm s} - 2 \omega_{\rm m}$, it is advantageous to linearize the operators using
\begin{subequations}\label{eq:linearization}
\begin{align}
    \hat{V}_{\rm in} =& \langle \hat{V}_{\rm in} \rangle + \hat{\delta V}_{\rm in},
    \\
    \hat{Q}_{\rm s} =& \langle \hat{Q}_{\rm s} \rangle +  \hat{\delta Q}_{\rm s},
    \\
    \hat{\Phi}_{\rm s} =& \langle \hat{\Phi}_{\rm s} \rangle + \hat{\delta \Phi}_{\rm s},
\end{align}
\end{subequations}
where
\begin{subequations}\label{eq:linearizationDef}
\begin{align}
    \langle \hat{V}_{\rm in} \rangle \equiv & \, A_{\rm in} e^{- i \omega_{\rm in} t} + \text{c.c.},
    \\
    \langle \hat{Q}_{\rm s} \rangle \equiv & \, 2 \sqrt{\hbar \omega_{\rm s} C_{\rm 0}} \alpha  \cos[\omega_{\rm in }t],
    \\
    \langle \hat{\Phi}_{\rm s} \rangle \equiv & \left(L_{\rm 0} + \frac{L}{2}\right) \frac{d \langle \hat{Q}_{\rm s} \rangle}{dt}  = \, - \sqrt{\frac{\hbar }{\omega_{\rm s} C_0}}\frac{\omega_{\rm in}}{\omega_{\rm s} } \alpha   \sin[\omega_{\rm in }t]
\end{align}.
\end{subequations}
Here $A_{\rm in}$ is the amplitude of the drive $\hat{V}_{\rm in}$ with $\gamma_{\rm t} =Z_{\rm out}/(L_{0} + L/2)$, and 
\begin{equation}\label{eq:alphaDef}
    \alpha = 4 \frac{A_{\rm in}}{(\omega_{\rm s}^2 - \omega_{\rm in}^2 - i \gamma_{\rm t} \omega_{\rm in})(L+ 2 L_{\rm 0})}.
\end{equation}
For convenience the phase of $A_{\rm in}$ is fixed such that $\alpha \in \mathbb{R}$.
Furthermore, $\hat{\delta V}_{\rm in}$ is the Johnson-Nyquist noise associated with the incoming field. To resemble a realistic experiment, we assume that the circuit is driven by a semi-infinite transmission line of impedance $Z_{\rm out}$ as in Ref.~\cite{dellantonio2018quantum}. The results in Sec.~\ref{sec:two_pho_cool} are derived assuming matching impedances between the circuit and the transmission line: $Z_{\rm out} = R_0 + R/2$. For generality, we keep $R_0$, $R$, and $Z_{\rm out}$ as distinct parameters in the following.

Plugging Eqs.~\eqref{eq:linearization} into Eqs.~\eqref{eq:FullHamiltonian}, we obtain the linearized Hamiltonian $\hat{\mathcal{H}}_{\rm lin}$
\begin{equation}\label{eq:H_linear_linquad_double_arm} 
\begin{split}
   \hat{\mathcal{H}}_{\rm lin} = & \hbar \omega_{\rm m} \hat{b}^\dagger \hat{b} + \frac{\hat{Q}^2_{\rm a}}{C_{\rm 0}} + \frac{\hat{\Phi}^2_{\rm a}}{4 L} +  \frac{ \hat{\delta Q}^2_{\rm s}}{4 C_{\rm 0}} + \frac{ \hat{\delta \Phi}^2_{\rm s} }{L + 2 L_{\rm 0}}  - x_{\rm zpm} \hat{F}_{\rm m} (\hat{b} + \hat{b}^\dagger)-  \hat{\delta Q}_{\rm s}(2\hat{\delta V}_{\rm in} + 2\hat{V}_{\rm R0} + \hat{V}_{ \rm R2}+ \hat{V}_{\rm R1} ) 
\\
&-
    2 \hat{Q}_{\rm a}(\hat{V}_{\rm R2} - \hat{V}_{\rm R1})  + \frac{g_1}{C_{\rm 0}\omega_{\rm s}} \hat{Q}_{\rm a}\langle\hat{Q}_{\rm s}\rangle (\hat{b} + \hat{b}^\dagger) + \frac{\delta g_{\rm 1}}{2 \omega_{\rm s} C_{\rm 0}}  \langle \hat{Q}_{\rm s} \rangle  \hat{\delta Q}_{\rm s} (\hat{b} + \hat{b}^\dagger) + \frac{g_{\rm 2}}{4 \omega_{\rm s} C_{\rm 0}}  \langle \hat{Q}_{\rm s} \rangle  \hat{\delta Q}_{\rm s} (\hat{b} + \hat{b}^\dagger)^2 
 \\
&+
 \underbrace{ \frac{g_1}{C_{\rm 0}\omega_{\rm s}} \hat{Q}_{\rm a}  \hat{\delta Q}_{\rm s} (\hat{b} + \hat{b}^\dagger) + \frac{\delta g_{\rm 1}}{4 \omega_{\rm s} C_{\rm 0}} \hat{\delta  Q}_{\rm s}^2 (\hat{b} + \hat{b}^\dagger) + \frac{\delta g_{\rm 1} }{\omega_{\rm s} C_{\rm 0}} \hat {Q}_{\rm a}^2 (\hat{b} + \hat{b}^\dagger)}_{\rm negligible}  \\
&+
 \underbrace{\frac{g_{\rm 2}}{8 \omega_{\rm s} C_{\rm 0}} \hat{\delta  Q}_{\rm s}^2 (\hat{b} + \hat{b}^\dagger)^2 + \frac{g_{\rm 2} }{2 \omega_{\rm s} C_{\rm 0}} \hat {Q}_{\rm a}^2 (\hat{b} + \hat{b}^\dagger) ^2}_{\rm negligible}  
 \\
 &  + \underbrace{\frac{\delta g_{\rm 1}}{2 \omega_{\rm s} 2 C_{\rm 0}} \langle\hat{Q}_{\rm s} \rangle^2(\hat{b} + \hat{b}^\dagger)}_{\text{rest position displacement}} + \underbrace{\frac{g_{\rm 2}}{4 \omega_{\rm s} 2 C_{\rm 0}} \langle\hat{Q}_{\rm s} \rangle^2(\hat{b} + \hat{b}^\dagger)^2}_{\text{mechanical frequency shift}},
\end{split}
\end{equation}
which can be used to determine the Heisenberg equations for the system operators. Before doing so, let us comment on the underlined terms in Eq.~\eqref{eq:H_linear_linquad_double_arm}. In the third and fourth rows, we collected all contributions that can be neglected, as they are not enhanced by the strong drive and/or are off-resonant. The two terms in the fifth row can be absorbed into the remainders by a shift in the rest position of the membrane \cite{MarquardtReview}, and a redefinition of the mechanical frequency $\omega_{\rm m}$.
The equations of motions resulting from the dynamically relevant terms in Eq.~\eqref{eq:H_linear_linquad_double_arm} are thus
\begin{subequations}\label{sys:double_arm_time}
\begin{align}
\begin{split}
    \dot{\hat{b}} =
    &
    \left( - i \omega_{\rm m} - \frac{\gamma_{\rm m}}{2} \right) \hat{b} + i \frac{x_{\rm zpm}}{\hbar} \hat{F}_{\rm m} - i \frac{g_1}{ \hbar C_{\rm 0}\omega_{\rm s}} \hat{Q}_{\rm a}\langle \hat{Q}_{\rm s} \rangle 
    \\
    & - i \frac{ \delta g_{\rm 1}}{2  \hbar \omega_{\rm s} C_{\rm 0}}\langle \hat{Q}_{\rm s} \rangle  \hat{\delta Q}_{\rm s} - i \frac{g_{\rm 2}}{2 \hbar \omega_{\rm s} C_{\rm 0}}\langle \hat{Q}_{\rm s} \rangle  \hat{\delta Q}_{\rm s} (\hat{b} + \hat{b}^\dagger) ,
    \label{sys:double_arm_time_a}
    \end{split}
    \\
    \dot{\hat{Q}}_{\rm a}=
    &
    \frac{\hat{\Phi}_a}{2 L}
    \label{sys:double_arm_time_b},
    \\
    \dot{\hat{\Phi}}_{\rm a}= 
    &
    - \frac{2 \hat{Q}_{\rm a}}{C_{\rm 0}} - \gamma_{\rm l} \hat{\Phi}_{\rm a}- \frac{g_1}{ C_{\rm 0}\omega_{\rm s}} \langle \hat{Q}_{\rm s}\rangle (\hat{b} + \hat{b}^\dagger) + 2 (\hat{V}_{\rm R2}- \hat V_{\rm R1}),
    \label{sys:double_arm_time_c}
    \\
    \dot{ \hat{ \delta Q}}_{\rm s}=
    &
    \frac{\hat{\delta \Phi}_{\rm s}}{  L_{\rm 0} + L/2},
    \label{sys:double_arm_time_d}
    \\
    \begin{split}
   \dot{\hat{\delta \Phi}}_{\rm s}= 
    &
    -  \frac{ \hat{\delta Q}_{\rm s}}{2 C_{\rm 0}} - \gamma \hat{\delta \Phi}_{\rm s} - \frac{\delta g_{\rm 1}}{2 C_{\rm 0} \omega_{\rm s}} \langle \hat{Q}_{\rm s}\rangle  (\hat{b} + \hat{b}^\dagger) - \frac{g_{\rm 2}}{4 C_{\rm 0} \omega_{\rm s}} \langle \hat{Q}_{\rm s}\rangle  (\hat{b} + \hat{b}^\dagger)^2 
    \\
    & +  (2 \hat{\delta V}_{\rm in} + 2 \hat V_{\rm R0} + \hat V_{\rm R2} + \hat V_{\rm R1}),
    \label{sys:double_arm_time_e}
    \end{split}
\end{align}
\end{subequations}
where we have included decays into Markovian reservoirs associated to each subsystem. Specifically, $\gamma_{\rm m}$ is the decay rate into the mechanical reservoir, $\gamma_{\rm l} = R/L$ is the decay rate into the reservoir of the asymmetric electrical mode, and $\gamma = \gamma_{\rm r} + \gamma_{\rm t}$ is the decay rate into the reservoirs (associated with $R_0+R/2$ and $Z_{\rm out}$, respectively) of the symmetric electrical mode. Here, $\gamma_{\rm r} = (R_{\rm 0} + R/2)/(L_{\rm 0}+L/2)$ and $\gamma_{\rm t}$ is defined above.

The system in Eqs.~\eqref{sys:double_arm_time} combines nonlinearly coupled differential equations, and does not have a simple analytical solution. However, under reasonable assumptions, it is possible to select the leading contributions to the dynamics -- which will lead us to the desired master equation -- and to remove negligible contributions. To do so, it is convenient to switch to the frequency domain. For an operator $\hat{O}(t)$, we define its Fourier series as
\begin{equation}\label{eq:FourierSeries}
 \hat{O}(t) = \sum_{i=-\infty}^{\infty} \hat{O}[\Omega_k] \frac{e^{-i \Omega_k t}}{\sqrt{\tau}},
\end{equation}
where $\Omega_k = 2\pi k/\tau$ $(k \in \mathbb{Z})$ are the allowed frequencies, and $\tau$ is a sufficiently long time interval. The Fourier coefficients $\hat{O}[\Omega_k]$ are then defined by
\begin{equation}\label{eq:FourierCoeff}
 \hat{O}[\Omega_k] = \int_{0}^{\tau} \hat{O}(t) \frac{e^{-i \Omega_k t}}{\sqrt{\tau}} dt.
\end{equation}
The differential equations in Eqs.~\eqref{sys:double_arm_time} can now be Fourier transformed into polynomial equations by using the property $(d \hat{O} / dt)[\Omega_k] = -i \Omega_k \hat{O}[\Omega_k]$. We can thus find relations for $\hat{Q}_{\rm a}$ and $\hat{\delta Q}_{\rm s}$: 
\begin{subequations}\label{sys:Fourier_charges}
\begin{align}
    \hat Q_{\rm a}[\Omega_k] =& \xi^{\gamma_{\rm l}}_{\omega_{\rm a}}[\Omega_k] \left[ - \sqrt{\hbar C_{\rm 0}} \frac{\alpha g_{\rm 1} \omega_{\rm a}^2}{2 \sqrt{\omega_{\rm s}}} \sum_{s = \pm 1}  \left(\hat b + \hat b^\dagger\right) [\Omega_k - s \, \omega_{\rm in}] + \left( \frac{\hat V_{\rm R2}[\Omega_k] - \hat V_{\rm R1}[\Omega_k]}{ L} \right)\right]
    \\
\begin{split}
    \hat{\delta Q}_{\rm s}[\Omega_k] = & 
    \xi^{\gamma}_{\omega_{\rm s}}[\Omega_k] \left[ - \sqrt{\hbar C_{\rm 0}}  \alpha \delta g_{\rm 1} \omega_{\rm s }^{\frac{3}{2}} \sum_{s = \pm 1} \left( \hat b + \hat b^\dagger \right) [\Omega_k - s \, \omega_{\rm in}]  + 2 \left(\frac{2 \hat{\delta V}[\Omega_k]+ 2 \hat V_{\rm R0}[\Omega_k] + \hat V_{\rm R2}[\Omega_k] + \hat V_{\rm R1}[\Omega_k]}{L + 2 L_{\rm 0}} \right) \right.
    \\
    &
   \left. \quad \qquad - \sqrt{\hbar C_{\rm 0}} \frac{\alpha g_{\rm 2} \omega_{\rm s }^{\frac{3}{2}}}{2} \sum_{\substack{s=\pm 1, \\ \Omega_l}} \left(  \hat b + \hat b^\dagger \right)[\Omega_k - s \, \omega_{\rm in} + \Omega_l] \left(\hat b + \hat b^\dagger\right)[- \Omega_l]   \right],
\end{split}
\end{align}
\end{subequations}
while Eq.~\eqref{sys:double_arm_time_a} for $\hat{b}$ becomes
\begin{equation}\label{eq:b_Fourier}
\begin{split}
    \left(i(\omega_{\rm m} - \Omega_k) +     \frac{\gamma_{\rm m}}{2}\right) \hat b[\Omega_k] =
    &
    i \frac{x_{\rm zpm}}{\hbar} \hat{F}_{\rm m} - i \frac{\alpha g_{\rm 1} }{\sqrt{\hbar \omega_{\rm s} C_{\rm 0}} } \sum_{s = \pm 1}  \hat Q_{\rm a}[\Omega_k - s \, \omega_{\rm in}]
    \\
    &
     - i \frac{\alpha \delta g_{\rm 1}}{2 \sqrt{\hbar \omega_{\rm s} C_{\rm 0}} }   \sum_{s = \pm 1} \hat{ \delta Q}_{\rm s}[\Omega_k - s \, \omega_{\rm in}]
     \\
     &
     - i \frac{\alpha g_2}{2 \sqrt{\hbar \omega_{\rm s} C_{\rm 0}} } \sum_{\substack{s=\pm 1, \\ \Omega_l}} \hat{\delta Q}_{\rm s}[\Omega_k - s \, \omega_{\rm in} + \Omega_l] \left( \hat b [- \Omega_l] + \hat b^\dagger [- \Omega_l] \right).
\end{split}
\end{equation}
Here, we have defined $\xi_{\rm \omega_0}^{\gamma}[\Omega_k] \equiv 1/(\omega_{\rm 0}^2 - \Omega_k^2 - i \gamma \Omega_k)$. 

Resorting to the Fourier transform does not in itself allow us to find solutions to the equations of motion in Eqs.~\eqref{sys:double_arm_time}. Indeed, operator products in the time domain become convolutions in the frequency domain, which can be taken care of only by considering infinite many terms. However, in the limit where the mechanical subsystem is characterized by a narrow bandwidth $\gamma_{\rm m}$, the frequency domain is useful for selecting the dominant dynamical terms \cite{dellantonio2018quantum}. In particular, it is reasonable to assume that the only relevant contributions in the sums in Eqs.~\eqref{sys:Fourier_charges} and \eqref{eq:b_Fourier} are the ones where the mechanical annihilation (creation) operator $\hat{b}$ ($\hat{b}^\dagger$) is centred at $+\omega_{\rm m}$ ($-\omega_{\rm m}$). After performing this frequency selection, we can insert Eqs.~\eqref{sys:Fourier_charges} into Eq.~\eqref{eq:b_Fourier} to obtain the solution for $\hat{b}[\Omega_k]$, and go back to the time domain to find
\begin{equation}\label{eq:b_time}
\begin{split}
    \dot{\hat b} = &
    \left( - i \omega_{\rm m} - \frac{\gamma_{\rm m}}{2}\right) \hat b + i \frac{x_{\rm zpm}}{\hbar} \hat F_{\rm m} + i \frac{\left( \alpha g_{\rm 1}\right)^2 \omega_{\rm a}^2}{2 \omega_{\rm s}} \left(\sum_{ s = \pm 1} \xi_{\rm \omega_{\rm a}}^{\gamma_{\rm l}}[\omega_{\rm m} - s\,  \omega_{\rm in}] \right) \hat b
    \\
    &
    +\left\lbrace \frac{-i\alpha g_{\rm 1}}{L\sqrt{\hbar \omega_{\rm s} C_{\rm 0}}} \sum_{s = \pm 1}  e^{-i \left( \omega_{\rm m} - s \, \omega_{\rm in} \right)t} \xi_{\omega_{\rm a}}^{\gamma_{\rm l}} [\omega_{\rm m} - s \, \omega_{\rm in}]\left(\hat V_{\rm R2} - \hat V_{\rm R1} \right)[\omega_{\rm m} - s \, \omega_{\rm in}]  + \text{H.c.} \right\rbrace
    \\
    &
    +\left\lbrace \frac{- i\delta g_{\rm 1}\omega_{\rm a}^2 }{ \hbar \omega_{\rm s}}  \sum_{ s = \pm 1} e^{-i \left( \omega_{\rm m} - s \, \omega_{\rm in} \right)t}  \xi_{\rm \omega_{\rm s}}^{\gamma}[\omega_{\rm m} - s \, \omega_{\rm in}] \left( 2\hat{\delta V}_{\rm in} + 2 \hat V_{\rm R0} + \hat V_{\rm R2} + \hat V_{\rm R1}\right)[\omega_{\rm m} - s \, \omega_{\rm in}] + \text{H.c.} \right\rbrace
    \\
    &
     + i \frac{\left(\alpha \delta g_{\rm 1}\right)^2 \omega_{\rm s}}{2 } \left( \sum_{ s = \pm 1} \xi_{\rm \omega_{\rm s}}^{\gamma}[\omega_{\rm m} - s\,  \omega_{\rm in}] \right) \hat b + i \frac{\left( \alpha g_{\rm 2}\right)^2 \omega_{\rm s}}{4}  \xi_{\omega_{\rm s}}^{\gamma} [\omega_{\rm s}] \hat b \hat b^\dagger \hat b 
    \\
    &
    - i \frac{\alpha g_{\rm 2} \omega_{\rm a}}{\sqrt{\hbar \omega_{\rm s} (L + 2 L_{\rm 0})}} e^{-i \omega_{\rm s} t}  \xi_{\rm \omega_{\rm s}}^{\gamma} [\omega_{\rm s}] \left( 2 \hat{\delta V}_{\rm in}+ 2 \hat V_{\rm R0} + \hat V_{\rm R2} + \hat V_{\rm R1}\right) [\omega_{\rm s}] \hat b^{\dagger}.
\end{split}
\end{equation}
Here, all noises need to be evaluated at the frequency indicated in the associated square brackets. This last equation describes the membrane's dynamics, \textit{including} the effects of the reservoirs of the electrical subsystems, but \textit{excluding} the symmetric and the asymmetric electrical modes. 
From Eq.~\eqref{eq:b_time}, it is possible to derive its corresponding effective Hamiltonian $\hat{\mathcal H}_{\rm eff}$ as
\begin{equation}\label{eq:eff_Hamiltonian}
\begin{split}
    \hat{\mathcal H}_{\rm eff} =& \hbar \omega_{\rm m} \hat b ^\dagger \hat b - x_{\rm zpm} \hat F_{\rm m}\left(\hat b + \hat b^\dagger \right) 
    \\
    &
    + \frac{\sqrt{\hbar}\alpha g_{\rm 1} \omega_{\rm a}}{\sqrt{L \omega_{\rm s}}} \left\lbrace \sum_{s = \pm 1} e^{-i\left( \omega_{\rm m} - s \,\omega_{\rm in} \right)t} \xi_{\rm \omega_{\rm a}}^{\gamma_{\rm l}}[\omega_{\rm m} - s \, \omega_{\rm in}] \left(\hat V_{\rm R2} - \hat V_{\rm R1} \right)[\omega_{\rm m} - s \, \omega_{\rm in}] \hat b^\dagger + \text{H.c.} \right\rbrace
    \\
     &
    + \frac{\sqrt{\hbar \omega_{\rm s}}\alpha \delta g_{\rm 1} }{ \sqrt{L+2L_0}} \left\lbrace\sum_{s = \pm 1} e^{-i\left( \omega_{\rm m} - s \,\omega_{\rm in} \right)t} \xi_{\rm \omega_{\rm s}}^{\gamma}[\omega_{\rm m} - s \, \omega_{\rm in}] \left(2 \hat{\delta V}_{\rm in} + 2 \hat V_{\rm R0} + \hat V_{\rm R2} + \hat V_{\rm R1} \right)[\omega_{\rm m} - s \, \omega_{\rm in}] \hat b^\dagger + \text{H.c.} \right\rbrace
    \\
    &
    + \frac{ \sqrt{\hbar \omega_{\rm s}} \alpha g_{\rm 2}}{2 \sqrt{L+2L_0}} \left\lbrace e^{-i \omega_{\rm s} t} \xi_{\rm \omega_{\rm s}}^{\gamma}[\omega_{\rm s}] \left(2 \hat{\delta V}_{\rm in}+ 2 \hat V_{\rm R0} + \hat V_{\rm R2} + \hat V_{\rm R1} \right)[\omega_{\rm s}] \hat b^\dagger \hat b^{\dagger} + \text{H.c.} \right\rbrace,
\end{split}
\end{equation}
which is also independent from the symmetric and asymmetric electrical modes. 

From $\hat{\mathcal H}_{\rm eff}$, a master equation \cite{walls2007quantum} that fully characterizes the membrane's dynamics can be derived. Under the assumptions described in Sec.~\ref{sec:two_pho_cool}, the correlators of the operators $\hat{\delta V}_{\rm in}$ and $\hat{V}_{Ri}$ ($i=0,1,2$) become
\begin{subequations}\label{Eq:VarianceElNoise}
	\begin{align}
	\left\langle \hat{\delta V}_{\rm in}\left[ \Omega_{h} \right]\hat{V}_{Ri}\left[ \Omega_{ m} \right] \right\rangle = & 0, \;\;\; \forall i , \\
	\left\langle \hat{V}_{Ri}\left[ \Omega_{h} \right]\hat{V}_{Rj}\left[ \Omega_{ m} \right] \right\rangle = & 0, \;\;\; \forall i\neq j, \\
	\left\langle \hat{\delta V}_{\rm in}\left[ \Omega_{h} \right]\hat{\delta V}_{\rm in}\left[ \Omega_{ m} \right] \right\rangle = & \frac{\hbar \Omega_{h} Z_{\rm out}}{2}\left[\bar{n}_{\rm e}(\Omega_{h},T_{\rm e})+ \theta\left(\Omega_{h}\right) \right]\delta\left(\Omega_{h}+\Omega_{ m}\right), \\
	\begin{split}
	\left\langle \hat{V}_{Ri}\left[ \Omega_{h} \right]\hat{V}_{Ri}\left[ \Omega_{ m} \right] \right\rangle = & \frac{\hbar \Omega_{h} R_i}{2}\left[\bar{n}_{\rm e}(\Omega_{h},T_{\rm e})+ \theta\left(\Omega_{h}\right) \right]\delta\left(\Omega_{h}+\Omega_{ m}\right),
	\end{split}
	\end{align}
\end{subequations}
with $R_1 = R_2 = R$ and $\bar{n}_{\rm e}(\Omega_{h},T_{\rm e})$ being the thermal occupation number of a reservoir at frequency $\Omega_{h}$ and temperature $T_{\rm e}$. With the correlators in Eqs.~\eqref{Eq:VarianceElNoise}, we derive the membrane's master equation in the standard Born-Markov approximation:
\begin{equation}\label{eq:MasterEq}
\begin{split}
    \dot{ \hat{ \rho}}_{\rm m} = 
    &
    \frac{1}{i \hbar} \left[\hat{\mathcal H}_{\rm m}, \, \hat \rho_{\rm m} \right] + ( \bar{n}_{\rm m} + 1) \frac{\gamma_{\rm m}}{2} \mathcal L[\hat b] \hat \rho_{\rm m} + \bar n_{\rm m} \frac{\gamma_{\rm m}}{2} \mathcal L[\hat{b}^\dagger; \hat{\rho}_{\rm m}]
    \\
    &
    +\gamma_{\rm l}\omega_{\rm a}^2 \left( \alpha g_{\rm 1}  \right)^ 2  \sum_{s = \pm 1} \left(\bar n_{\rm e} + \frac{1 + s}{2} \right) \left\lvert \xi_{\rm \omega_{\rm a}}^{\gamma_{\rm l}} [\omega_{\rm m} + s \, \omega_{\rm in}] \right\rvert ^2 \mathcal{L}[\hat{b}; \hat{\rho}_{\rm m}]
    \\
    &
    +\gamma_{\rm l}\omega_{\rm a}^2 \left( \alpha g_{\rm 1}  \right)^ 2  \sum_{s = \pm 1} \left(\bar n_{\rm e} + \frac{1 + s}{2} \right) \left\lvert \xi_{\rm \omega_{\rm a}}^{\gamma_{\rm l}} [\omega_{\rm m} - s \, \omega_{\rm in}] \right\rvert^2 \mathcal{L}[\hat{b}^\dagger; \hat{\rho}_{\rm m}]
    \\
    &
    +\gamma \omega_{\rm s}^2 \left( \alpha \delta g_{\rm 1} \right)^ 2   \sum_{s = \pm 1} \left(\bar n_{\rm e} + \frac{1 + s}{2} \right) \left\lvert \xi_{\rm \omega_{\rm s}}^{\gamma} [\omega_{\rm m} + s \, \omega_{\rm in}] \right\rvert^2 \mathcal{L}[\hat{b}; \hat{\rho}_{\rm m}]
    \\
    &
    +\gamma \omega_{\rm s}^2 \left( \alpha \delta g_{\rm 1} \right)^ 2 \sum_{s = \pm 1} \left(\bar n_{\rm e} + \frac{1 + s}{2} \right) \left\lvert \xi_{\rm \omega_{\rm s}}^{\gamma} [\omega_{\rm m} - s \, \omega_{\rm in}] \right\rvert^2 \mathcal{L}[\hat{b}^\dagger; \hat{\rho}_{\rm m}]
    \\
    &
    +\frac{\gamma \omega_{\rm s}^2 \left( \alpha g_{\rm 2} \right)^ 2}{4} \left\lvert \xi_{\rm \omega_{\rm s}}^{\gamma} [\omega_{\rm s}] \right\rvert ^ 2 (\bar n_{\rm e} + 1) \mathcal{L}[\hat{b} \hat{b}; \hat{\rho}_{\rm m}] ,
\end{split}
\end{equation}
where $\bar{n}_{\rm m}$ is the average occupation of the mechanical reservoir. For a generic operator $\hat{O}$ the Lindblad dissipator is defined as $\mathcal{L}[\hat{O}; \hat{\rho}_{\rm m}] = 2 \hat{O} \hat{\rho}_{\rm m} \hat{O}^\dagger - \acomm{\hat{O}^\dagger \hat{O}}{\hat{\rho}_{\rm m}}$. For simplicity, we set the thermal photonic occupations of all electrical reservoirs to be approximately the same, and equal to $\bar{n}_{\rm e}$. 

From the master equation, it is possible to directly obtain the rates $\Gamma_2$, $\Gamma_1^{\rm (r)}$ and $\Gamma_1^{\rm (b)}$ defined in Sec.~\ref{sec:two_pho_cool}.
The form given there is derived with the approximations:
\begin{subequations}
\begin{align}
    \left\lvert\xi_{\rm \omega_{\rm a}}^{\gamma_{\rm l}}[\omega_{\rm m} \pm \omega_{\rm in}] \right\rvert ^ 2 \simeq & \, \frac{1}{\omega_{\rm a}^4},
    \\
    \left\lvert\xi_{\rm \omega_{\rm s}}^{\gamma}[\omega_{\rm m} \pm \omega_{\rm in}] \right\rvert ^ 2 \simeq & \, \frac{1}{5 \mp 4} \frac{1}{4 \omega_{\rm m}^2 \omega_{\rm s}^2} ,
    \\
    \bar n_{\rm e } \simeq & \, 0, \label{eq:assumptionTemp}
\end{align}
\end{subequations}
which are satisfied under the considered assumptions $\omega_{\rm a} \gg \omega_{\rm s} \gg \omega_{\rm m} \gg \gamma $. A cryogenic temperature is required in order for Eq.~\eqref{eq:assumptionTemp} to be valid for microwave frequencies (for the considered parameters $\bar n_{\rm e } \simeq 0.1$ at $140$ mK). The linear heating and cooling processes associated to $\Gamma_1^{\rm (b)}$ can be identified in the second and third rows of Eq.~\eqref{eq:MasterEq}, and originate from the term $\propto g_1 \hat{Q}_{\rm a} \hat{Q}_{\rm s}(\hat{b} + \hat{b}^{\dagger})$ in Eq.~\eqref{eq:FullHamiltonian}. The processes associated to $\Gamma_1^{\rm (r)}$ and attributed to the term $\propto \delta g_1 \hat{Q}_{\rm s}^2(\hat{b} + \hat{b}^{\dagger})$ in Eq.~\eqref{eq:FullHamiltonian} appear on the fourth and fifth rows in Eq.~\eqref{eq:MasterEq}. Finally, the two-phonon cooling, from the non-Gaussian interaction in Eq.~\eqref{eq:FullHamiltonian}, is found in the last row of the master equation, while the first row contains coherent dynamics and processes associated to the intrinsic mechanical reservoir.

\section{Mechanical beam splitter interaction}\label{sec:drivingField}
To coherently manipulate the mechanical state, in Sec.~\ref{sec:qntm_man} we have employed a beam splitter Hamiltonian of the form
\begin{equation}\label{eq:bs_ham}
    \hat{\mathcal{H}}_{\rm bs} = \hbar \frac{\Omega \hat{b} + \Omega^{*}\hat{b}^\dagger}{2}
    .
\end{equation}
In this section, we explain how to experimentally implement such interaction by using two input fields $\hat{V}_1$ and $\hat{V}_2$ at frequencies $\omega_1$ and $\omega_2$ such that $\omega_1 - \omega_2 = \omega_{\rm m}$. We assume that both $\hat{V}_1$ and $\hat{V}_2$ are off-resonant with all other system dynamics, so that it does not interfere with the two-phonon cooling. 

The Hamiltonian considered here is of the form $\hat{\mathcal{H}} = \hat{\mathcal{H}}_{\rm m} + \hat{\mathcal{H}}_{\rm o} + \hat{\mathcal{H}}_{\rm int}$, where $\hat{\mathcal{H}}_{\rm m} = \hbar \omega_{\rm m} \hat{b}^\dagger \hat{b}$ is the mechanical term considered throughout this work, and
\begin{subequations}
\begin{align}
    \hat{\mathcal{H}}_{\rm o} & = 
    \frac{\hat{\Phi}_{\rm o}^{2}}{2L_{\rm o}} + \frac{\hat{Q}_{\rm o}^{2} }{2 C_{\rm o}} - 2 \hat{Q}_{\rm o} (\hat{V}_1 + \hat{V}_2)
    , \\
    \hat{\mathcal{H}}_{\rm int} & = 
    \tilde{g}_{\rm{1}} \frac{\hat{Q}_{\rm{o}}^2}{2 \omega_{\rm{o}} C_{\rm{o}}} (\hat{b} + \hat{b}^\dagger)
    . \label{eq:int_ham_om}
\end{align}
\end{subequations}
Here, subscript ``o'' is used to indicate an electrical mode with intrinsic inductance and capacitance $L_{\rm o}$ and $C_{\rm o}$, respectively, that is coupled to the membrane via the linear coefficient $\tilde{g}_1$. For practical purposes, this electrical mode can be the symmetric mode, such that $L_{\rm o} \rightarrow L_0 + L/2$, $C_{\rm o} \rightarrow 2C_0$ and exploiting the inevitable fabrication imperfections we set $\tilde{g}_1 \rightarrow \delta g_1$. In this case, via the superposition principle, we can analyse the effects from the drivings $\hat{V}_1$ and $\hat{V}_2$ independently from the other dynamics studied in Sec.~\ref{sec:two_pho_cool} and in App.~\ref{sec:mast_eq_der}, and include them afterwards in the effective Hamiltonian $\hat{\mathcal{H}}_{\rm bs}$ in Eq.~\eqref{eq:bs_ham}.

Following the same steps as in App.~\ref{sec:mast_eq_der}, we assume that the input fields $\hat{V}_1$ and $\hat{V}_2$ are strongly driven with averages given by $\langle \hat{V}_{i}\rangle = A_{i} e^{- i \omega_{i} t} + c.c.$ ($i=1,2$). As a consequence, the average charge becomes
\begin{equation}\label{eq:avg_el_o}
    \langle \hat{Q}_{\rm{o}} \rangle = \sqrt{\frac{\hbar \omega_{\rm{o}} C_{\rm{o}}}{2}} (\alpha_1 e^{i \omega_{1} t} + \alpha_2 e^{i \omega_{2} t} + c.c.)
    ,
\end{equation}
where $\alpha_i$ ($i=1,2$) is defined by $\alpha_i = 2 A_{i} /[(\omega_{\rm o}^2 - \omega_{i}^2 - i \gamma_{\rm o} \omega_{i})L_{\rm o}]$ [see also Eq.~\eqref{eq:alphaDef}]. Here, $\gamma_{\rm o}$ is the decay rate of the mode ``o'' into the transmission line used to drive the cavity with the input fields $\hat{V}_1$ and $\hat{V}_2$. We remark that, in principle, the drives considered here also affect the other electrical modes, such as the symmetric and asymmetric studied in the rest of this manuscript. In practice, these contributions can be made negligible by tuning the frequencies $\omega_1$ and $\omega_2$.

With the purpose of deriving the beam splitter Hamiltonian in Eq.~\eqref{eq:bs_ham} we use the mean field approximation for the charge $\hat{Q}_{\rm o}$. By plugging Eq.~\eqref{eq:avg_el_o} into Eq.~\eqref{eq:int_ham_om} we find that the mechanical subsystem is subjected to an effective interaction of the form
\begin{equation}\label{eq:bs_ham_der}
     \hat{\mathcal{H}}_{\rm int} 
     \rightarrow 
     \frac{\hbar \tilde{g}_{\rm{1}}}{4} ( \alpha_{\rm{1}} \alpha_{\rm{2}}^\ast e^{i (\omega_{\rm{1}} - \omega_{\rm{2}})t} \hat{b} + h.c.).
\end{equation}
In the interaction picture, for $\omega_1 - \omega_2 = \omega_{\rm m}$ the time dependence $e^{i (\omega_{\rm{1}} - \omega_{\rm{2}})t}$ is compensated by the free evolution of the mechanical annihilation operator $\hat{b}$. Therefore, upon the substitution $\Omega = \tilde{g}_1 \alpha_1 \alpha_{2}^{*}/2$, Eq.~\eqref{eq:bs_ham_der} is the beam splitter Hamiltonian. This means that from the membrane's point of view, the dynamics introduced by the two fields $\hat{V_1}$ and $\hat{V}_2$ driving the electrical mode ``o'' can be viewed as the desired $\hat{\mathcal{H}}_{\rm bs}$ used in the main text for the coherent manipulation.

We note that exploiting the fabrication imperfection to achieve a linear coupling $\tilde{g}_1 \rightarrow \delta g_1$ is perfectly compatible with the general requirement of a negligible $\delta g_1$ to prevent linear heating. The undesired heating is proportional to the vacuum fluctuations and suppressed by being far off resonance. In contrast the linear coupling is a resonant process enhanced by the driving field making is much stronger for the same asymmetry parameter. A similar argument can be made for the linear cooling assumed for initialization of the system before the coherent drive, although in this case the enhancement only comes from being on resonance.

\section{Dynamical contributions from the residual linear couplings}\label{sec:dynmc_cntrbtns}
\begin{figure}
    \centering
    \includegraphics[width=0.75\linewidth]{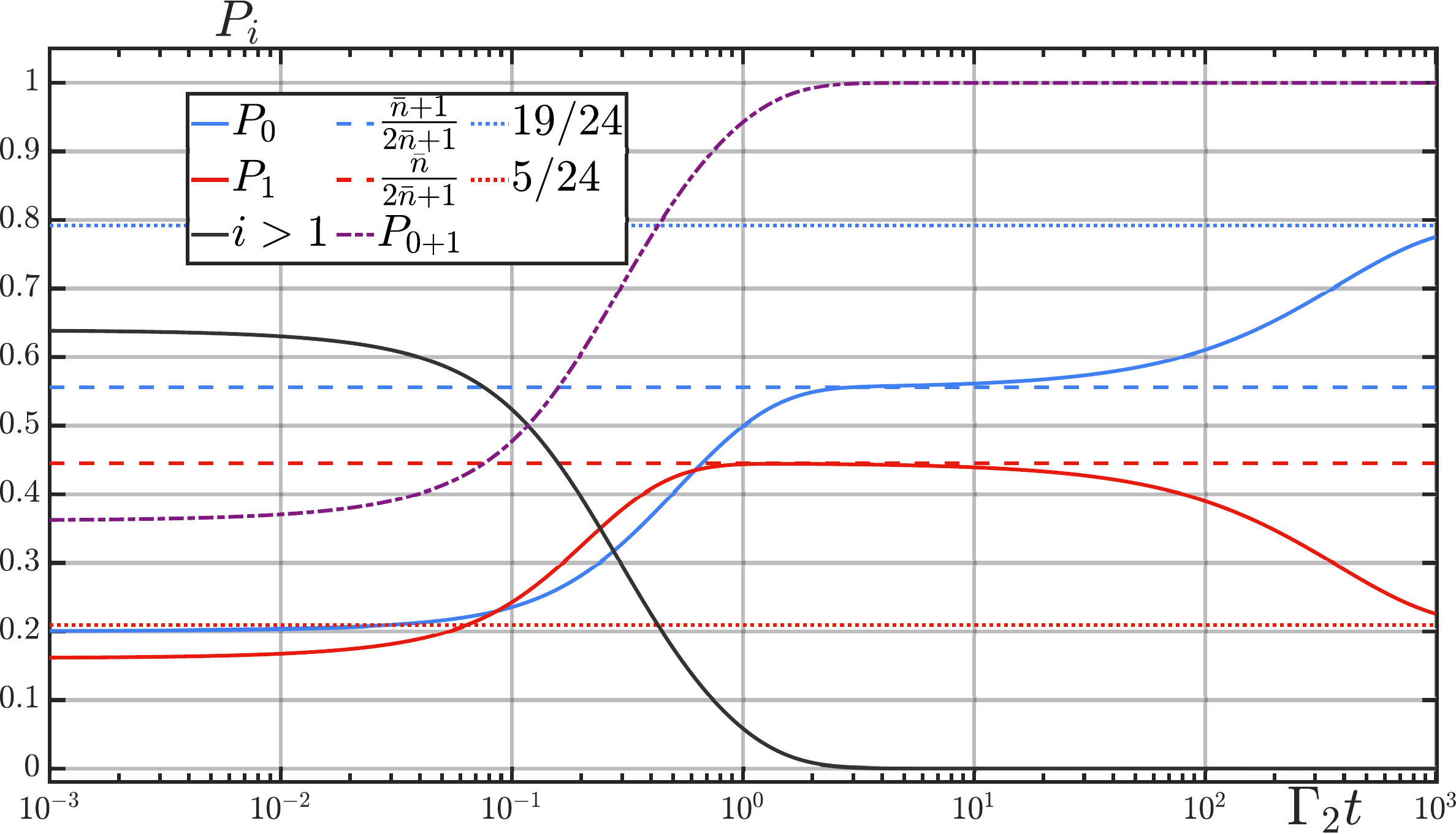}
    \caption{
    Time evolution of the mechanical Fock states' populations $P_i = \langle i \rvert \hat{\rho}_{\rm m}\lvert i \rangle$ (full lines) with the two-phonon cooling active. We assume $\lambda = 2000$ and initialize the system in a thermal state with an average of $\bar{n} = 4$ phonons. The black line represents the sum $\sum_{j>1}\langle j \rvert \hat{\rho}_{\rm m}\lvert j \rangle$, and the purple dotted-dashed represent $P_{0+1}=P_0+P_1$. Dashed lines are the expected populations of a perfect two-phonon cooling process (see Sec.~\ref{sec:expt_cons}), which is achieved transiently when $1 \lesssim \Gamma_2 t\lesssim 10^2$. For longer times, detrimental linear processes kick in and drive the system into its steady state, as highlighted by the dotted lines. All curves are obtained by simulating Eq.~\eqref{eq:master_equation}; the parameters used are the same as in Fig.~\ref{fig:plot_twopho_cool}.
    }
    \label{fig:supplemental}
\end{figure}

In this section, we analyze the effects of residual linear couplings on the system's dynamics. We consider the same settings as in Secs.~\ref{sec:two_pho_cool} and \ref{sec:expt_cons}. In the interesting regime $\lambda \gg 1$, the two-phonon cooling is dominant and the system promptly relaxes into its two lowest energy states. This can be seen in Fig.~\ref{fig:plot_twopho_cool}, from which it is evident that in a timescale $t \sim \Gamma_2^{-1}$ we obtain the desired outcome of a perfect two-phonon cooling process. Here, we investigate the membrane's dynamics afterwards, i.e., on a timescale $t \gtrsim \lambda/\Gamma_2$ in which detrimental heating effects disturb the mechanical two-level system's coherence.

The resulting dynamics on longer time scales are shown in Fig.~\ref{fig:supplemental}. In this plot, we use the same parameters as in Fig.~\ref{fig:plot_twopho_cool}, but simulate the system for longer times to demonstrate the detrimental effects of the linear couplings. Since the two-phonon cooling is always the dominant process, it is possible to neglect all Fock states except the ground and first excited mechanical  states. The effective dynamics is then characterized by the system of Eqs.~\eqref{sys:rhodot_2times2} with $\Omega = 0$. By setting the time derivatives to zero, the steady states are found to be
\begin{equation}\label{eq:steady_detrimental}
    P_1 = \langle 1 \rvert \hat{\rho}_{\rm m}\lvert 1 \rangle \overset{t\rightarrow \infty}{\longrightarrow} \frac{1}{4}\frac{\Gamma_1^{\rm (r)} + 9\Gamma_1^{\rm (b)}}{3\Gamma_1^{\rm (r)}+9\Gamma_1^{\rm (b)}},
\end{equation}
and $P_0 = 1 - P_1$. In case $\Gamma_1^{\rm (r)} = \Gamma_1^{\rm (b)}$, we obtain $P_1$ and $P_2$ to be $5/24$ and $19/24$, respectively, which are the dotted lines in Fig.~\ref{fig:supplemental}. Full lines are obtained by numerically simulating Eq.~\eqref{eq:master_equation}, which yields the exact membrane's dynamics, and we include the purple dotted-dashed line, representing the total population $P_{0+1} = P_0 + P_1$ of the two lowest mechanical states. 

As it is possible to see in Fig.~\ref{fig:supplemental}, as long as the two-phonon cooling is active, there is no leakeage outside the subspace of the ground and first excited states. This can be understood by looking at the purple dashed-dotted line that reaches its asymptote $1$ for $t \gtrsim \Gamma_2^{-1}$. However, while the $\lbrace \lvert 0 \rangle, \lvert 1 \rangle \rbrace$ subspace is always protected by the two-phonon cooling, the coherence within is still limited by detrimental linear effects. In fact, on a timescale $t \propto (\Gamma_1^{\rm (r)} + \Gamma_1^{\rm (b)})^{-1}$, the membrane reaches a thermal state within the protected two-level sector with populations described by Eq.~\eqref{eq:steady_detrimental}.

\section{Time evolution of pure states}\label{sec:pureStatesEvolution}

In Secs.~\ref{sec:two_pho_cool} and \ref{sec:expt_cons} we considered the mechanical state before the two-phonon cooling to be  a thermal state. This choice is motivated by it resembling realistic experimental scenarios. At the same time the density matrices of thermal states are diagonal, making the numerical computation simpler. This allowed us to push our simulations to include more states, increase the precision and to run for longer periods of time.

In this section, we investigate the opposite scenario. Namely, we consider an initial highly coherent mechanical state whose density matrix has a large off-diagonal component. First, we demonstrate that the steady state of a perfect two-phonon cooling process is such that $\hat{\rho}_{ij} = 0$ for all $i,j \neq 0,1$, i.e., all entries of the mechanical density matrix $\hat{\rho}_{\rm m}$ outside the lowest two levels' subspace are zero. Then, we consider the initial mechanical state
\begin{equation}\label{eq:psi_k}
    \vert \Psi_k \rangle = \frac{1}{\sqrt{2k}} \sum_{i=0}^{2k-1} \vert i \rangle,
\end{equation}
and show that a perfect two-phonon cooling process applied to it results in a mixed state that retains large coherences $\hat{\rho}_{01}$ and $\hat{\rho}_{10}$ between $\ket{0}$ and $\ket{1}$.

To prove that a perfect two-phonon cooling process asymptotically (for $t\rightarrow \infty$) eliminates all entries $\hat{\rho}_{ij}$ ($i,j \neq 0,1$) of $\hat{\rho}_{\rm m}$, we start from the master equation in Eq.~\eqref{eq:master_equation}. By setting $\gamma_{\rm m} = \Gamma_{1}^{\rm (r)} = \Gamma_{1}^{\rm (b)} = 0$, we find the relation $\dot{\hat{\rho}}_{\rm m} = \Gamma_2 \mathcal{L}[\hat{b} \hat{b};\hat{\rho}_{\rm m}]/4$, that in the steady state $\dot{\hat{\rho}}_{\rm m} = 0$ entails $\hat{\rho}_{ij} = 0$ for all $i,j \neq 0,1$. This demonstrates that not only the diagonal, but also all the off-diagonal elements of the mechanical density matrix outside the two lowest energy levels subspace, asymptotically reach zero when the two phonon cooling is active. 

The question is then which state within the $\lbrace \lvert 0 \rangle, \lvert 1 \rangle \rbrace$ subspace the membrane reaches after a perfect two-phonon cooling process. This depends on the initial mechanical state. As we have seen in Sec.~\ref{sec:expt_cons}, a thermal state will be projected onto $\left[(\bar{n} + 1) \lvert 0 \rangle \langle 0 \rvert + \bar{n} \lvert 1 \rangle \langle 1 \rvert \right]/(2\bar{n}+1)$, that has no off-diagonal elements, with  $\bar{n}$ being the initial thermal occupation. If we use $\vert \Psi_k \rangle$ in Eq.~\eqref{eq:psi_k} instead, it is possible to prove that the resulting two-level density matrix $\hat{\rho}_{\rm m}$ has the following elements
\begin{subequations}
\begin{align}
    \hat{\rho}_{00} = \hat{\rho}_{11} = &  \frac{1}{2}
    \label{eq:gen_plus_cool_diag}, \\
    \hat{\rho}_{01} = \hat{\rho}_{10} = &  \frac{1}{2k} \sum_{i=0}^{k-1} \frac{1}{4^{i}} \binom{2i}{i} \sqrt{2i+1} \xrightarrow{k\to \infty} \frac{1}{\sqrt{2\pi}}
    \label{eq:gen_plus_cool_offdiag}.
\end{align}
\end{subequations}
Qualitatively, Eq.~\eqref{eq:gen_plus_cool_diag} follows from symmetry and normalization. Since the initial state $\vert \Psi_k \rangle$ in Eq.~\eqref{eq:psi_k} has equal contributions from even and odd Fock states, the mechanical density matrix at the end of the cooling must have equal populations in its two lowest energy levels. Determining the off diagonal elements $\hat{\rho}_{01}$ and $\hat{\rho}_{10}$ is more complicated. With Eq.~\eqref{eq:gen_plus_cool_offdiag} derived by induction, the reason is that only fractions of the off-diagonal elements of the initial density   are transferred into $\hat{\rho}_{01}$ and $\hat{\rho}_{10}$ and the rest are lost. Specifically, these fractions depend on the decay rates of the two Fock states $\ket{i}$ and $ \ket{j}$ involved in the decay of $\ket{i}\bra{j}$. More energetic states (with higher $i$ and $j$) are associated with a larger loss. This difference in decay rates reduces the coherence between the states, i.e. the rates at which states decay provide ``which-way'' information about the state, which disturbs the coherence. This effect is enhanced with more decays and as a result $\hat{\rho}_{01}$ and $\hat{\rho}_{10}$ in Eq.~\eqref{eq:gen_plus_cool_offdiag} are monotonically decreasing functions of the parameter $k$. 

\begin{figure}
    \centering
    \includegraphics[width=0.75\linewidth]{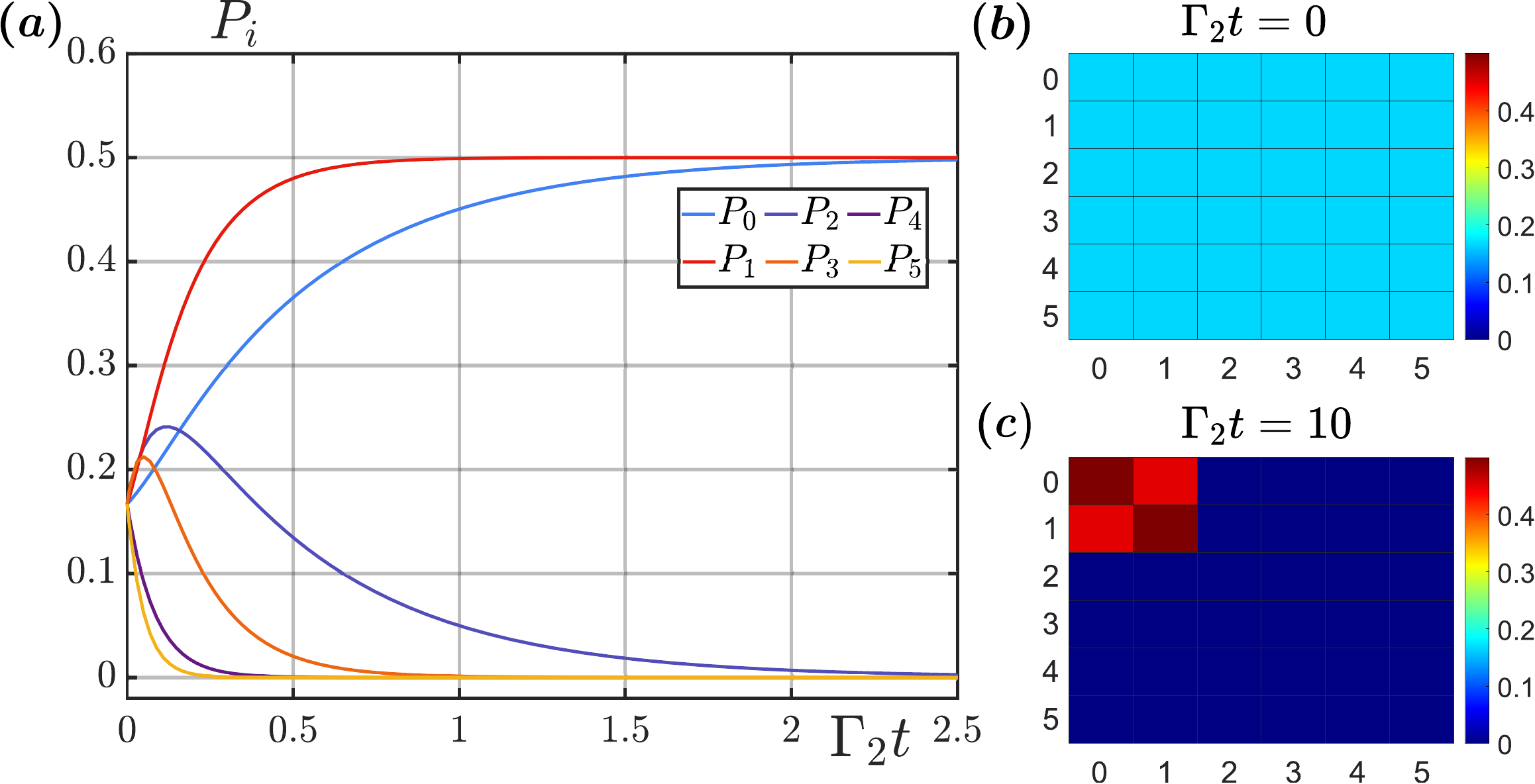}
    \caption{
    In \textit{(a)}, we present the time evolution of the mechanical Fock states' populations $P_i = \langle i \rvert \hat{\rho}_{\rm m}\lvert i \rangle$ (full lines) with the two-phonon cooling active. We assume $\lambda = 2000$ and initialize the system in the superposition state $\vert \Psi_3 \rangle = (\vert 0 \rangle + \vert 1 \rangle + \vert 2 \rangle + \vert 3 \rangle + \vert 4 \rangle + \vert 5 \rangle)/\sqrt{6}$. The time $\Gamma_2 t = 2.5$ is where $P_0 \simeq P_1$.
    In \textit{(b)} and \textit{(c)}, we present the density matrices at $t=0$ and at $\Gamma_2 t = 10$
    respectively. These panels demonstrate how the two-phonon cooling concentrates the mechanical state into the two lowest energy levels. Furthermore, starting from an initial coherent state, it prepares a coherent superposition in the $\{\ket{0},\ket{1} \}$ subspace as evidenced by the off-diagonal elements of the density matrix. All curves are derived by simulating Eq.~\eqref{eq:master_equation}; the parameters used are the same as in Figs.~\ref{fig:plot_twopho_cool} and \ref{fig:supplemental}.
    }
    \label{fig:Fig2s}
\end{figure}

Albeit this last fact, as proven by the asymptotic limit for $k \to \infty$ in Eq.~\eqref{eq:gen_plus_cool_offdiag}, even for very large values of $k$ the coherence of the initial state survives to a large degree. For $k=3$ we find that $\hat{\rho}_{01} = \hat{\rho}_{10} \simeq 0.451$, which we confirm by  numerical simultion. In Fig.~\ref{fig:Fig2s} we present a simulation of the whole system with $\vert \Psi_3 \rangle$ [see Eq.~\eqref{eq:psi_k}] as initial state and the two-phonon cooling being the dominant process. As it is possible to see in panel (a), the probabilities of being in each Fock state (full lines) are reduced to zero from $1/6$, except for $P_0$ and $P_1$ that both reach approximately the value in Eq.~\eqref{eq:gen_plus_cool_diag}. This is further demonstrated by the density matrices depicted in panels (b) and (c), describing the system at the initial and final times, respectively. Specifically, the latter is compatible with the values of $\hat{\rho}_{01}$ or $\hat{\rho}_{10}$ in Eq.~\eqref{eq:gen_plus_cool_offdiag}, in agreement with the above discussion. 


%
%


\section{Quadratic coupling vs strong coupling regime}\label{sec:optical_regimes}

An important question to be addressed is whether the system reaches any strong coupling regime simultaneously with our non-Gaussian operations from the quadratic coupling. This would imply that the Master equation in Eq.~\eqref{eq:master_equation}
would not be valid. To address this concern, we would like to make three distinctions about relevant kinds of strong couplings. First, there is the (non-Gaussian) single photon strong coupling regime, characterized by $g_1 \gg \gamma$ ($\gamma$ being the decay rate of the main electrical mode, which is assumed to be more damped than the mechanical one) \cite{YanbeiChen}. Achieving this regime is known to be extremely challenging. We refer to \cite{dellantonio2018quantum} for demonstrating that this limit is far more demanding than $\lambda \gg 1$ (also this regime is driving independent). Second, there is the (Gaussian) linear strong coupling regime, which can be obtained – if the coupling is resonant – when $g_1 \alpha \gg \gamma$ \cite{Marquardt2008Nature}. In our settings, the linear coupling is suppressed both by symmetry and because the main electrical mode is far off resonance. As long as $\lambda \gg 1$, the effect of the linear coupling is thus smaller than that of the quadratic, hence we would actually reach the quadratic strong coupling regime \textit{before} the linear one. Finally, the quadratic strong coupling regime is characterized by $g_2 \alpha \gg \gamma$. To be dominated by the two-photon cooling as opposed to the linear heating, we need $g_2^2 \lvert \alpha \rvert^2/\gamma \gg \gamma_{\rm m} (2 \bar{n}_{\rm m}+1)$. The combination of these two requirements  again puts a requirement on the quality factor $Q$ of the mechanical oscillator. The parameters considered in this work are $Q\in [10^5,10^7]$, $\gamma/(2\pi)=150$ kHz and $\gamma_{\rm m} (2 \bar{n}_{\rm m}+1)\leq 1$ kHz (see Sec.~\ref{sec:expt_cons}). Hence, there is ample margin to adjust the driving strength such that
\begin{equation}
   \gamma \gg \frac{g_2^2 \lvert \alpha \rvert^2}{\gamma} \gg \gamma_{\rm m} (2 \bar{n}_{\rm m}+1),
\end{equation}
as required for our master equation in Eq.~\eqref{eq:master_equation}
to work.  Therefore, we believe that the quadratic interaction considered by us is a very promising route to show non-Gaussian effects in electro-mechanics, and the analysis we provided is accurate in the assumed parameter regimes. This is also supported by numerical simulations of the composite system before tracing out the electrical symmetric and asymmetric modes [Eqs.~\eqref{sys:double_arm_time} in App.~\ref{sec:mast_eq_der}]. 

\end{document}